\documentclass[sn-mathphys]{sn-jnl}

\usepackage{amsfonts,amsmath,mathrsfs,amssymb}
\usepackage{amsthm}%
\DeclareMathOperator*{\argmax}{arg\,max}

\usepackage{graphics}
\usepackage{graphicx}
\usepackage{subfigure}
\usepackage{multirow}%
\usepackage{mathrsfs}%
\usepackage{xcolor}%
\usepackage{textcomp}%
\usepackage{manyfoot}%
\usepackage{booktabs}%
\usepackage{algorithm}%
\usepackage{algorithmicx}%
\usepackage{algpseudocode}%
\usepackage{listings}%
\usepackage{threeparttable}

\theoremstyle{thmstyleone}%
\newtheorem{theorem}{Theorem}
\theoremstyle{thmstyletwo}%
\newtheorem{remark}{Remark}%

\theoremstyle{thmstylethree}%

\begin{document}
\title{Doubly robust estimation of optimal treatment regimes for survival data using an instrumental variable}

\author[1,2]{\fnm{Xia} \sur{Junwen}}
\equalcont{These authors contributed equally to this work.}

\author[3]{\fnm{Zhan} \sur{Zishu}}
\equalcont{These authors contributed equally to this work.}

\author*[1,2]{\fnm{Zhang} \sur{Jingxiao}}\email{zhjxiaoruc@163.com}

\affil[1]{\orgdiv{Center for Applied Statistics}, \orgname{Renmin University of China}, \orgaddress{\city{Beijing}, \postcode{100872}, \country{China}}}

\affil[2]{\orgdiv{School of Statistics}, \orgname{Renmin University of China}, \orgaddress{\city{Beijing}, \postcode{100872}, \country{China}}}

\affil[3]{\orgdiv{School of Public Health}, \orgname{Southern Medical University}, \orgaddress{\city{Guangzhou}, \postcode{510515}, \country{China}}}

\abstract{
 In survival contexts, substantial literature exists on estimating optimal treatment regimes, where treatments are assigned based on personal characteristics to maximize the survival probability. These methods assume that a set of covariates is sufficient to deconfound the treatment-outcome relationship. However, this assumption can be limited in observational studies or randomized trials in which non-adherence occurs. Therefore, we propose a novel approach to estimating optimal treatment regimes when certain confounders are unobservable and a binary instrumental variable is available. Specifically, via a binary instrumental variable, we propose a semiparametric estimator for optimal treatment regimes by maximizing a Kaplan-Meier-like estimator of the survival function. Furthermore, to increase resistance to model misspecification, we construct novel doubly robust estimators. Since the estimators of the survival function are jagged, we incorporate kernel smoothing methods to improve performance. Under appropriate regularity conditions, the asymptotic properties are rigorously established. Moreover, the finite sample performance is evaluated through simulation studies. Finally, we illustrate our method using data from the National Cancer Institute's prostate, lung, colorectal, and ovarian cancer screening trial.
}


\maketitle

\section{Introduction}
In clinical settings, assigning the same treatment to all patients may be suboptimal because different subgroups may benefit from different treatments. The goal of optimal treatment regimes is to assign treatments based on individual characteristics, thereby achieving maximum clinical outcomes or so-called values \cite[]{murphy2003optimal,zhao2012estimating,zhang2012a,laber2015treebased,shi2018highdimensional}. 
For chronic diseases such as cancer, survival data often arises, 
where the outcome of interest, the survival time, is subject to right censoring. 
Previous works for developing optimal treatment regimes for survival data were built on a critical assumption that no unmeasured confounder exists so that the relationship between the treatment and outcome can be deconfounded by observed covariates \cite[]{gengOptimalTreatmentRegimes2015,baiOptimalTreatmentRegimes2017,jiangEstimationOptimalTreatment2017,zhouTransformationInvariantLearningOptimal2022}. However, the assumption can not be guaranteed in observational studies or randomized trials with non-adherence issues. 
For example, in an intervention experiment designed to assess the effect of a screening test or a drug, relatively healthy individuals may be more likely to skip the interventions \cite[]{kianianCausalProportionalHazards2021,leeDoublyRobustNonparametric2021}. However, they are also more likely to have better clinical outcomes, making health status an unmeasured confounder. 

To address this limitation, an alternative variable called an instrumental variable (IV) has been utilized to estimate optimal treatment regimes for completely observed outcomes. An IV is a pretreatment variable that is independent of all unmeasured confounders and exerts its effect on outcomes solely through treatments. Common IVs include assignments in randomized trials with possible non-adherence,
genetic variants known to be associated with the phenotypes, and calendar periods as a determinant of patients' access to treatments \cite[]{leeDoublyRobustNonparametric2021,yingTwostageResidualInclusion2019,mackCalendarTimeSpecificPropensity2013,tchetgentchetgenInstrumentalVariableEstimation2015}.
The use of an IV to estimate optimal treatment regimes began with \cite{cuiSemiparametricInstrumentalVariable2021}, who considered the problem from a classification perspective. Later, \cite{qiuOptimalIndividualizedDecision2021} studied the problem through stochastic regimes. Both assumed that an IV can be used to point identify the treatment effect under modified assumptions introduced by \cite{wangBoundedEfficientMultiply2018}. When an IV can only partially identify the treatment effect, suggesting that only a lower bound and an upper bound of the treatment effect can be obtained, \cite{puEstimatingOptimalTreatment2021} and \cite{chenEstimatingImprovingDynamic2021} proposed to maximize the lower bound of the treatment effect to obtain harmless optimal treatment regimes under different data structures. However, existing studies are not suitable for survival data since they can not handle the challenge introduced by right censoring. In this article, with the help of an IV, we aim to 
identify the treatment effect and estimate optimal treatment regimes under survival data. 

The motivating example of this work is the prostate, lung, colorectal, and ovarian (PLCO) cancer screening trial \cite[]{teamProstateLungColorectal2000}, a two-arm randomized trial evaluating screening tests for PLCO cancers. Between November 1993 and July 2001, ten centers across the U.S. recruited participants and collected the data through 2015. We will focus on determining the optimal assignment of flexible sigmoidoscopy screening (a screening test for colorectal cancer) to increase individual life expectancy.
The outcome of interest, survival time, is subject to right censoring due to reasons such as the limited time of the trial.
It is important to personalize the flexible sigmoidoscopy screening because not all individuals will experience the desired benefit \cite[]{tangTimeBenefitColorectal2015}.
Additionally, there is non-adherence in the trial, which could be influenced by a range of unmeasured confounders \cite[]{leeDoublyRobustNonparametric2021,kianianCausalProportionalHazards2021}. For example, relatively healthy individuals may be more likely to skip the screening. When unmeasured confounders are present, the estimated optimal treatment regimes based on the assumption of their absence become unreliable. 
Overall, the PLCO data involves the two challenges we mentioned before: First, some patients are censored; Second, the no unmeasured confounding assumption is violated.
Fortunately, the assigned treatment as an IV can be employed to compensate for the bias caused by unmeasured confounding. 

This article makes a number of contributions to both the IV and optimal treatment regime literature. First, we introduce an inverse weighted Kaplan-Meier estimator with an IV (IWKME-IV) of the counterfactual survival function (other survival-related value functions are also allowed)
and a corresponding semiparametric estimator of optimal treatment regimes by maximizing the IWKME-IV.
Our method allows the use of an IV to overcome unmeasured confounding.
In addition, we propose their doubly robust versions to increase resistance to model misspecification. 
To the best of our knowledge, there is no prior literature on the doubly robust estimator for estimating the treatment effect using an IV in survival contexts, not to mention assigning treatments.
Previous doubly robust estimators are based on the assumption of no unmeasured confounding \cite[]{jiangEstimationOptimalTreatment2017} or focus on the treatment effect among compliers \cite[]{leeDoublyRobustNonparametric2021}. Compared to the doubly robust estimator in \cite{wangBoundedEfficientMultiply2018,cuiSemiparametricInstrumentalVariable2021}, which considered the non-censored data with an IV, our doubly robust estimator can account for potential bias introduced by a misspecified censoring model. Second, we consider the smoothing approach to improving the performance of the estimators. The value function is non-smoothed with respect to the parameters of the treatment regimes, which leads to computational challenges in the optimization. The kernel smoothing technique solves the challenges efficiently and maintains consistency. Third, we prove theoretical guarantees for our proposed estimators. Specifically, we provide asymptotic bounds for our estimators, which highlights the advantages of the doubly robust estimator: it is not only consistent if one of the models is correctly specified, but can also incorporate some semiparametric models or nonparametric models to have a root-\(n\) consistency.

The remaining sections of this article are organized as follows. Section \ref{method} presents the mathematical foundations and estimators for utilizing an IV to determine optimal treatment regimes, including the doubly robust and kernel-smoothed estimators. {Section \ref{sec3} provides the asymptotic properties of these estimators. The finite sample performance is explored through simulations in Section \ref{4}. Section \ref{5} illustrates the application using the PLCO dataset. Appendix \hyperref[app B]{A} discusses the IV-related assumptions used in this article. Appendix \hyperref[intuition]{B} provides some intuition for our estimator.} Proofs and additional
numerical studies can be found in the supporting information. In addition, our proposed method is implemented using R, and the R package \texttt{otrKM} to reproduce our results is available at \url{https://cran.r-project.org/web/packages/otrKM/index.html}.


\section{Methodology} \label{method}
Consider a binary treatment indicator \( A \in\{0, 1\} \), an unmeasured confounder (possibly a vector-value) \(U\), a binary IV \(Z \in \{0,1\}\), and a set of fully observed covariates \(\boldsymbol{L}\) (a vector with \(p\) dimensions). An individual treament regime is a function map \(d(\cdot )\in\mathcal{D}\) from the patient's baseline covariates \(\boldsymbol{L}\) to the treatment \(0\) or \(1\) such that the patient with baseline covariates \(\boldsymbol{L}\) would receive the treatment \(0\) if \(d(\boldsymbol{L})=0\) or the treatment \(1\) if \(d(\boldsymbol{L})=1\). We may apply machine learning methods to estimate optimal treatment regimes, such as the neural networks \cite[]{mi2019bagging} and the random forest \cite[]{doubleday2018an}, which correspond to \(\mathcal{D}\) indexed by a significant number of parameters. However, it is also important to consider regimes with interpretability and transparency, especially in clinical medicine. Thus, we assume \( \mathcal{D}=\{d_{\boldsymbol{\eta}}: d_{\boldsymbol{\eta}}\left(\boldsymbol{L}\right)= I\{\tilde{\boldsymbol{L}}^{T}{\boldsymbol{\eta}}  \geq 0 \}, \|{\boldsymbol{\eta}}\|_2=1 \} \) in this article, where \(\tilde{\boldsymbol{L}}=(1,\boldsymbol{L}^{T})^{T}\) and \(\|{\boldsymbol{\eta}}\|_2=\sqrt{\sum_{i=1}^{p+1}\eta_i^2}\).
Let \(T\) be the continuous survival time of interest, with the conditional survival function \(S_T(t|Z,\boldsymbol{L}, A)=P(T \geq t|Z,\boldsymbol{L},A)\) and the corresponding conditional cumulative hazard function \(\Lambda_T(t|Z,\boldsymbol{L},A)=-\mathrm{log}\{S_T(t|Z,\boldsymbol{L},A)\}\). Due to limited time and refusals to answer, the survival time \(T\) may not be observable; in such cases, the censoring time \(C\) replaces \(T\). Thus, the observation is \(\tilde{T} = \text{min}\{T,C\}\) and \(\delta=I\{T\leq C \}\). In this article, we assume that the complete data \( \{(U_{i},\boldsymbol{L}_{i}, A_{i}, Z_{i}, \tilde{T}_{i}, \delta_{i}), i=1, \ldots, n \} \) are independent and identically distributed across \( i \).

The counting process is defined as \( N(t) = I\{\tilde{T} \leq t, \delta=1\} \), and the at-risk process is denoted by \( Y(t) = I\{\tilde{T} \geq t\} \). 
The counting process of the censoring time is defined as \( N_C(t) = I\{\tilde{T} \leq t, \delta=0\} \); the at-risk process of the censoring time is denoted by \( Y_C(t) = I\{\tilde{T} \geq t\} \). It is notable that \(Y_C(t)=Y(t)\), and we will alternatively use \(Y(t)\) and \(Y_C(t)\) for clarity. Counterfactual results are denoted with a superscript star. Given treatment \(A=a\), let \(T^*(a)\) denote the counterfactual survival time, \(N^{*}(a ; t)=I\left[\min \left\{T^{*}(a), C\right\} \leq\right. \left.t, T^{*}(a) \leq C\right] \) denote the counterfactual counting process, and \( Y^{*}(a ; t)=I\left[\min \left\{T^{*}(a), C\right\} \geq t\right] \) denote the counterfactual at-risk process. Under the regime \(d_{\boldsymbol{\eta}}\), define the counterfactual survival time as \(T^{*}\left(d_{\boldsymbol{\eta}}(\boldsymbol{L})\right) = T^*(1) I\{d_{\boldsymbol{\eta}}(\boldsymbol{L})=1\}+T^*(0) I\{d_{\boldsymbol{\eta}}(\boldsymbol{L})=0\} \). The counterfactual counting process 

and the counterfactual at-risk process under the regime \(d_{\boldsymbol{\eta}}\) are defined accordingly.

In the following, for any function \(f\) of \(d_{\boldsymbol{\eta}}(\boldsymbol{L})\), its definition is \(f(d_{\boldsymbol{\eta}}(\boldsymbol{L}))=\sum_{a\in\{0,1\}}f(A=a)I\{d_{\boldsymbol{\eta}}(\boldsymbol{L})=a\}\). We can also utilize the counterfactual censoring time \(C^*(a)\) to define the counterfactual counting process and the at-risk process. These two definitions will produce identical estimators with slightly different assumptions. Further details and explanations can be found in Remark \ref{remark1}.

In the context of optimal treatment regimes, the counterfactual survival function under regime \(d_{\boldsymbol{\eta}}\), \(S^{*}(t; \boldsymbol{\eta}) = P\left[T^{*}\left(d_{\boldsymbol{\eta}}(\boldsymbol{L})\right)>t\right]\), is of interest. For a predetermined \(t\), \(S^{*}(t; \boldsymbol{\eta})\) as a value function provides a basis for the definition of optimal treatment regimes.
\begin{equation*}
   d_{{{\boldsymbol{\eta}}^{opt}}}(\boldsymbol{L}) = \argmax _{\mathcal{D}} S^{*}(t ; {\boldsymbol{\eta}})=I(\tilde{\boldsymbol{L}}^{T}{\boldsymbol{\eta}}^{opt}  \geq 0 ),
\end{equation*}
where 
\begin{equation*}
  {\boldsymbol{\eta}}^{opt}=\argmax _{\|\boldsymbol{\eta}\|_2=1} S^{*}(t ; \boldsymbol{\eta}).
\end{equation*}

In this article, despite the optimality of regimes being defined by the largest \(t\)-year survival probability, other value functions, such as the restricted mean \cite[]{gengOptimalTreatmentRegimes2015} and the quantile \cite[]{zhouTransformationInvariantLearningOptimal2022} of the survival time, are also available. We defer the discussion to Section \ref{discussion}.

\subsection{An estimator of the counterfactual survival function with unmeasured confounders}
A consistent estimator for \(S^{*}(t; \boldsymbol{\eta})\) is necessary to estimate optimal treatment regimes. The uninformative censoring assumption is usually utilized to describe the correlation between the survival time and the censoring time. That is, the counterfactual survival time is conditionally independent of the censoring time given the observed covariates as follows:

\vspace{1ex}
\textbf{A1:} (Uninformative censoring). \( T^{*}(a) \perp \! \! \! \perp C | Z, \boldsymbol{L}, A \) for \(a=0,1\).
\vspace{1ex}

Let \( S_{C}\{s | Z, \boldsymbol{L}, A\}=P(C \geq s | Z,\boldsymbol{L},A)\) denote the conditional survival function of the censoring time and \(\Lambda_C(t|Z,\boldsymbol{L},A)=-\mathrm{log}\{S_C(t|Z,\boldsymbol{L},A)\}\) denote the corresponding conditional cumulative hazard function of the censoring time. If the \(d_{\boldsymbol{\eta}}\)-specific counterfactual counting process \( N_{i}^{*}\left(d_{\boldsymbol{\eta}}(\boldsymbol{L}_i) ; s\right) \) and the counterfactual at-risk process \( Y_{i}^{*}\left(d_{\boldsymbol{\eta}}(\boldsymbol{L}_i); s\right) \) for all $0\leq s \leq t$ were observed, an intuitive estimator of \( S^{*}(t; \boldsymbol{\eta}) \) could be the inverse probability of censoring weighted Kaplan-Meier estimator. 
\begin{equation} \label{S}
  \widehat{S}^{*}(t ; \boldsymbol{\eta})=\prod_{s \leq t}\left\{1-\frac{\sum_{i=1}^{n}d N_{i}^{*}\left(d_{\boldsymbol{\eta}}(\boldsymbol{L}_i) ; s\right)/S_{C}\{s | Z_i, \boldsymbol{L}_i, d_{\boldsymbol{\eta}}(\boldsymbol{L}_i)\} }{\sum_{i=1}^{n}Y_{i}^{*}\left(d_{\boldsymbol{\eta}}(\boldsymbol{L}_i) ; s\right)/S_{C}\{s | Z_i, \boldsymbol{L}_i, d_{\boldsymbol{\eta}}(\boldsymbol{L}_i)\} }\right\}
\end{equation}



In clinical studies, since the counterfactual outcomes \(T^*_i(0)\) and \(T^*_i(1)\) used to calculate \( N_{i}^{*}\left(d_{\boldsymbol{\eta}}(\boldsymbol{L}_i); s\right) \) and \( Y_{i}^{*}\left(d_{\boldsymbol{\eta}}(\boldsymbol{L}_i); s\right) \) are not observable at the same time, \( \widehat{S}^{*}(t; \boldsymbol{\eta}) \) is not computable with the observed data. In order to obtain a proper estimator based on the observed data, tremendous literature has been established on the assumption of no unmeasured confounding \cite[]{gengOptimalTreatmentRegimes2015,baiOptimalTreatmentRegimes2017,jiangEstimationOptimalTreatment2017,zhouTransformationInvariantLearningOptimal2022}.

\vspace{1ex}
\textbf{A2\(^*\):} (No unmeasured confounding). \( T^*(a) \perp \! \! \! \perp A | \boldsymbol{L} \) for \( a=0, 1 \).
\vspace{1ex}

However, assumption A2 is too restrictive to be applied in observational studies or in randomized trials in which non-adherence occurs. Its violation will lead to an inaccurate estimate of the optimal treatment regime.
To address the problem, a ubiquitous method in economics and epidemiology called IV analysis \cite[]{wangIVEstimationCausal2022,cuiSemiparametricInstrumentalVariable2021,qiuOptimalIndividualizedDecision2021,wangBoundedEfficientMultiply2018} can be utilized to construct an estimator for \(S^{*}(t; \boldsymbol{\eta})\). Write the counterfactual survival time under the IV \(Z=z\) and treatment \(A=a\) as \(T^*(z, a)\); let \(f(Z | \boldsymbol{L})=P(Z|\boldsymbol{L})\) be the conditional probability mass function of \(Z\) given \(\boldsymbol{L}\), \(\pi(A | Z,\boldsymbol{L})=P(A|Z,\boldsymbol{L},U)\) be the conditional probability mass function of \(A\) given \(Z\), \(\boldsymbol{L}\) and \(U\); define \(\delta(\boldsymbol{L}) = \pi(A=1| Z=1, \boldsymbol{L})-\pi(A=1 | Z=0,\boldsymbol{L})\) and \(\widetilde{\delta}(\boldsymbol{L}, U) = \pi(A=1 | Z=1, \boldsymbol{L},U)-\pi(A=1 | Z=0, \boldsymbol{L},U)\). An IV requires the following assumptions.

\vspace{1ex}
\textbf{A2:} (Latent unconfoundedness). \( T^*(z,a) \perp \! \! \! \perp Z,A | \boldsymbol{L},U \) for \( z,a=0, 1 \).

\textbf{A3:} (IV relevance). \( Z \not \! \perp \! \! \! \perp A | \boldsymbol{L} \).

\textbf{A4:} (Exclusion restriction). \( T^*(z, a)=T^*(a) \) for \( z, a=0, 1 \) almost surely.

\textbf{A5:} (IV independence). \( Z \perp \! \! \! \perp U | \boldsymbol{L} \).

\textbf{A6:} (IV positivity). \( 0<f(Z=1 | \boldsymbol{L})<1 \) almost surely.

\textbf{A7:} (Independent adherence type). \(\delta(\boldsymbol{L}) =\widetilde{\delta}(\boldsymbol{L}, U)\)  almost surely.
\vspace{1ex}


\begin{figure}[!htbp]
    \centering
    \includegraphics[width=0.4\textwidth]{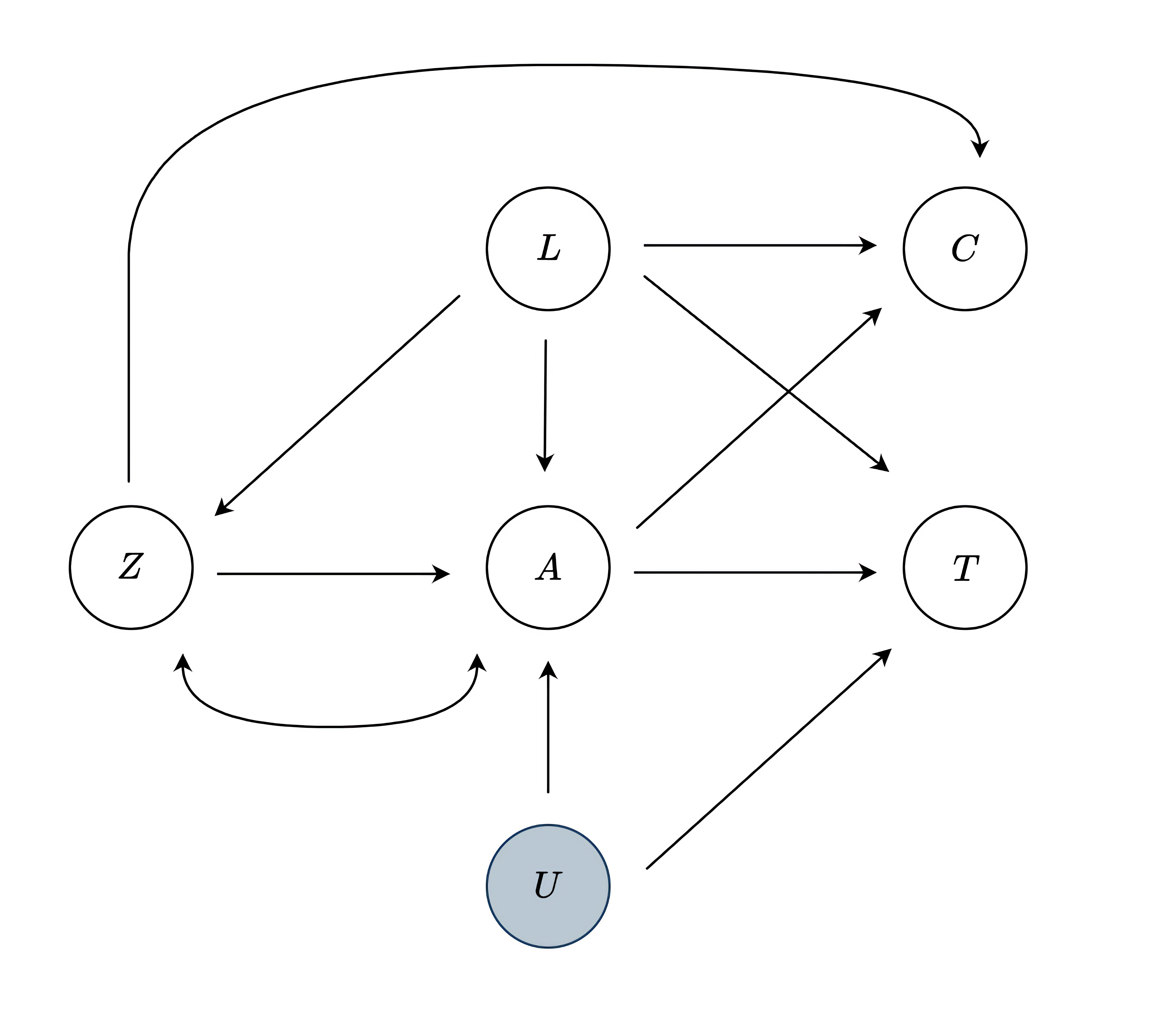}
    \caption{Causal graph representing the instrumental variable model defined by assumptions \(\mathrm{A}1\)-\(\mathrm{A}7\). The directed arrow from, for example, \(Z\) to \(A\) indicates that \(Z\) is a cause of \(A\). The bi-directed arrow between \(Z\) and \(A\) indicates potential unmeasured common causes of \(Z\) and \(A\). The independence or conditional independence among variables can be observed from the graph via the Markov assumption \cite[]{pearl1995causal}.}
    \label{fig:causal graph}
\end{figure}

Figure \ref{fig:causal graph} gives an illustration of the IV assumptions. In our case, no unmeasured confounding assumption is problematic as health status is a confounder. However, the IV assumptions are more likely to be accepted. More detailed discussion on these IV-related assumptions and why our case satisfies these assumptions can be found in Appendix \hyperref[app B]{A}.

With assumptions A\(1\)-A\(7\), the IWKME-IV is as follows. {The intuition underlying the IWKME-IV can be found in Appendix \hyperref[intuition]{B}.}
\begin{equation}\label{S_I}
    \widehat{S}_{I}^*(t ;\boldsymbol{\eta})=\prod_{s \leq t}\left\{1-\frac{\sum_{i=1}^{n} {\widehat{\omega}_i}(d_{\boldsymbol{\eta}}(\boldsymbol{L}_i);s) d N_{i}(s)}{\sum_{i=1}^{n} {\widehat{\omega}_i}(d_{\boldsymbol{\eta}}(\boldsymbol{L}_i);s) Y_{i}(s)}\right\},
\end{equation}
where 
\[\widehat{\omega}_i(d_{\boldsymbol{\eta}}(\boldsymbol{L}_i);s)= \frac{(2Z_i-1) (2A_i-1) I\{A_i=d_{\boldsymbol{\eta}}(\boldsymbol{L}_i)\}}{\widehat{\delta}(\boldsymbol{L}_i) \widehat{f}(Z_i | \boldsymbol{L}_i)\widehat{S}_{C}\{s | Z_i, \boldsymbol{L}_i, d_{\boldsymbol{\eta}}(\boldsymbol{L}_i)\}}. \]

Here, \(\widehat{\delta}(\boldsymbol{L}_i)=\widehat{\pi}(A_i=1| Z_i=1, \boldsymbol{L}_i)-\widehat{\pi}(A_i=1 | Z_i=0,\boldsymbol{L})\), \(\widehat{f}(Z_i|\boldsymbol{L}_i)\), and \(\widehat{S}_{C}\{s | Z_i, \boldsymbol{L}_i, d_{\boldsymbol{\eta}}(\boldsymbol{L}_i)\}\) are estimators of the nuisance functions \(\delta(\boldsymbol{L}_i)\), \(f(Z_i|\boldsymbol{L}_i)\), and \(S_C\{s | Z_i, \boldsymbol{L}_i, d_{\boldsymbol{\eta}}(\boldsymbol{L}_i)\}\).
We can posit, for example, the logistic model (parametric models), the single-index model (semiparametric models), and the random forest (nonparametric models) to estimate \(\pi(A| Z, \boldsymbol{L})\) and \(f(Z|\boldsymbol{L})\), thus obtaining \(\widehat{\delta}( \boldsymbol{L}_i)\) and \(\widehat{f}(Z_i|\boldsymbol{L}_i)\).
For \(\widehat{S}_C\{s | Z_i, \boldsymbol{L}_i, d_{\boldsymbol{\eta}}(\boldsymbol{L}_i)\}\), we can posit nonparametric models such as the random survival forest \cite[]{leeDoublyRobustNonparametric2021} and the local Kaplan-Meier estimator \cite[]{gonzalez-manteigaAsymptoticPropertiesGeneralized1994}, as well as semiparametric models that leverage working models such as the additive proportional hazards model \cite[]{leeDoublyRobustNonparametric2021} and the Cox proportional hazards model \cite[]{zhaoDoublyRobustLearning2015} to estimate \(S_C\{s|Z,\boldsymbol{L}, A\}\) and therefore obtain \(\widehat{S}_C\{s | Z_i, \boldsymbol{L}_i, d_{\boldsymbol{\eta}}(\boldsymbol{L}_i)\}\).

\vspace{1em}
\begin{remark} \label{remark1}

    In the aforementioned analysis, we define the counterfactual counting process and the at-risk process without considering the counterfactual censoring time \(C^*(a)\). Additionally, we introduce the assumption \textnormal{A\(1\)} to account for the dependence between the censoring time and treatment. 

    Another way to introduce the dependence is to use \(C^*(a)\) when defining the counterfactual counting process and the at-risk process, i.e. \(Y^*(a;t)=I[\min\{T^*(a),C^*(a)\}\geq t]\) and \(N^{*}(a ; t)=I\left[\min \left\{T^{*}(a), C^*(a)\right\} \leq\right. \left.t, T^{*}(a) \leq C^*(a)\right] \). In such a case, our estimator (\ref{S_I}) and the following doubly robust estimator (\ref{S_D}) are still consistent with a modified assumption \textnormal{A1\(^*\)} alternative to the assumption \textnormal{A1} \cite[]{baiOptimalTreatmentRegimes2017,wangIVEstimationCausal2022}. Proofs are left in the supporting information.
    
\vspace{1ex}
\textnormal{\textbf{A1\(^*\):}} \( T^{*}(a) \perp \! \! \! \perp C^*(a) | Z, \boldsymbol{L}, A \) and \( C^{*}(a) \perp \! \! \! \perp A | Z, \boldsymbol{L} \) for \(a=0,1\).
\vspace{1ex}
\end{remark}

\begin{remark}
When administrative censoring occurs such that \( S_{C}\{s | Z_i, \boldsymbol{L}_i, A_i\}=P(C_i \geq s)\), \(\widehat{\omega}_i(d_{\boldsymbol{\eta}}(\boldsymbol{L}_i);s)\) can be simplified as follows.
\[\widehat{\omega}_i(d_{\boldsymbol{\eta}}(\boldsymbol{L}_i);s)= \frac{(2Z_i-1) (2A_i-1) I\{A_i=d_{\boldsymbol{\eta}}(\boldsymbol{L}_i)\}}{\widehat{\delta}(\boldsymbol{L}_i) \widehat{f}(Z_i | \boldsymbol{L}_i)}. \]
\end{remark}

\subsection{A doubly robust estimator of the counterfactual survival function}
The IWKME-IV (\ref{S_I}) may produce unreliable results if the models for $f(Z | \boldsymbol{L})$ and \(S_{C}\{s | Z, \boldsymbol{L}, A\}\) are not properly specified. Moreover, as demonstrated in Section \ref{sec3}, the estimator is inefficient (a lower convergence rate than root-\(n\)) when positing semiparametric models for the nuisance models. Hence, a more desirable estimator would possess both the doubly robust property and efficiency by incorporating assumed model information. For example, one strategy could involve positing a Cox proportional hazards model for the conditional survival function \(S_{T}(t | Z, \boldsymbol{L}, A)\).
Other semiparametric models (such as the additive hazards model) and nonparametric models (such as the random survival forest) are also allowable.

{
A popular method to obtain the doubly robust estimator is through the efficient influence function (EIF), which yields an asymptotically optimal estimator \cite[]{tchetgentchetgenInstrumentalVariableEstimation2015,wangBoundedEfficientMultiply2018,tsiatisSemiparametricTheoryMissing2006}. However,
the derivation of the EIF for the IWKME-IV is challenging.
First, it is a product limit estimator; second, the survival time and the censoring time are partially observed; and third, in the IV setting, the correlation among variables is complex. 
Consequently, we endeavor to derive novel EIFs for \(E\{ \omega(d_{\boldsymbol{\eta}}(\boldsymbol{L});s) dN(s)\}\) and \(E \{\omega(d_{\boldsymbol{\eta}}(\boldsymbol{L});s) Y(s)\}\) with the Gateaux derivative technique \cite[]{hines2022demystifying}. The Gateaux derivative can be used to compute the directional derivative of the estimand, such as \(E\{ \omega(d_{\boldsymbol{\eta}}(\boldsymbol{L});s) dN(s)\}\), along the path of all parametric submodels of the nuisance functions. The parametric submodels of the nuisance functions, such as \(f(Z|L)\), are \(f_\epsilon(Z|L)=(1-\epsilon)f(Z|L)+\epsilon\tilde{f}(Z|L)\), where \(\epsilon\in [0,1]\) and \(\tilde{f}(Z|L)\) is a deterministic distribution used to disturb \(f(Z|L)\). The Gateaux derivative at point \(\epsilon=0\) is the EIF. The efficient estimators based on the EIFs can be subsequently established by utilizing the Taylor expansion (or so-called von Mises expansion) with the basis of the Gateaux derivative. The incorporation of derivative information is the reason behind the appealing properties of the efficient estimators. The EIFs for \(E\{ \omega(d_{\boldsymbol{\eta}}(\boldsymbol{L});s) dN(s)\}\) and \(E \{\omega(d_{\boldsymbol{\eta}}(\boldsymbol{L});s) Y(s)\}\) are placed in the supporting information. 
}

With the EIFs, we construct a novel doubly robust Kaplan-Meier estimator with an IV (DRKME-IV) as follows, where \( 1/n\sum_{i=1}^n \widehat{\psi}_{l,i}(d_{\boldsymbol{\eta}}(\boldsymbol{L}_i);s)\), \(l=1,2\) are efficient estimators for \(E\{ \omega(d_{\boldsymbol{\eta}}(\boldsymbol{L});s) dN(s)\}\) and \(E \{\omega(d_{\boldsymbol{\eta}}(\boldsymbol{L});s) Y(s)\}\), respectively. To our knowledge, there is no prior literature on the doubly robust estimator for estimating the treatment effect using an IV in survival contexts.

\begin{equation}\label{S_D}
  \widehat{S}_{D}^*(t ;\boldsymbol{\eta})=\prod_{s \leq t}\left\{1-\frac{\sum_{i=1}^{n} \widehat{\psi}_{1,i}(d_{\boldsymbol{\eta}}(\boldsymbol{L}_i);s)}{\sum_{i=1}^{n}\widehat{\psi}_{2,i}(d_{\boldsymbol{\eta}}(\boldsymbol{L}_i);s)}\right\},
\end{equation}

where
\begin{align*}
  \widehat{\psi}_{l,i}(d_{\boldsymbol{\eta}}(\boldsymbol{L}_i);s)=&\frac{2Z_i-1}{\widehat{\delta}(\boldsymbol{L}_i) \widehat{f}(Z_i | \boldsymbol{L}_i)}\Big\{\frac{(2A_i-1) g_l(\tilde{T}_i,\delta_i; s) I\{A_i=d_{\boldsymbol{\eta}}(\boldsymbol{L}_i)\}}{\widehat{S}_{C}\{s | Z_i, \boldsymbol{L}_i, d_{\boldsymbol{\eta}}(\boldsymbol{L}_i)\}}\\
  &-\widehat{\gamma}^{'}_l(\boldsymbol{L}_i,d_{\boldsymbol{\eta}}(\boldsymbol{L}_i) ;s)-[A_i-\widehat{\pi}(A_i=1 | Z_i=0, \boldsymbol{L}_i)]\widehat{\gamma}_l(\boldsymbol{L}_i,d_{\boldsymbol{\eta}}(\boldsymbol{L}_i);s)\Big\}+\widehat{\gamma}_l(\boldsymbol{L}_i,d_{\boldsymbol{\eta}}(\boldsymbol{L}_i);s)\\
  &+\frac{(2Z_i-1) (2A_i-1) I\{A_i=d_{\boldsymbol{\eta}}(\boldsymbol{L}_i)\}}{\widehat{\delta}(\boldsymbol{L}_i) \widehat{f}(Z_i | \boldsymbol{L}_i)}\int_{0}^{s^-} \widehat{E}\{m_l(T_i; s)|T_i\geq r, Z_i, \boldsymbol{L}_i, d_{\boldsymbol{\eta}}(\boldsymbol{L}_i) \} \\
  &\times \frac{d \widehat{M}_C\{r| Z_i, \boldsymbol{L}_i, d_{\boldsymbol{\eta}}(\boldsymbol{L}_i)\}}{\widehat{S}_{C}\{r^+ | Z_i, \boldsymbol{L}_i, d_{\boldsymbol{\eta}}(\boldsymbol{L}_i)\}}, \ l=1,2,\ i=1,...,n.
\end{align*}

Here, for a function \(f\) of \(x\), \(f(a^-)\) and \(f(a^+)\) denote the left limit and the right limit of \(f(x)\) at point \(a\), respectively.
The functions \(g_l(\tilde{T}_i,\delta_i;s)\) and \(m_l(T_i;s)\) are defined as \(g_1(\tilde{T}_i,\delta_i;s)=dN_i(s)\), \(g_2(\tilde{T}_i,\delta_i; s)=Y_i(s)\), \(m_1(T_i;s)=dI\{T_i \leq s\}\), and \(m_2(T_i;s)=I\{T_i \geq s\}\). The explicit forms of \(\widehat{\gamma}_l(\boldsymbol{L}_i,d_{\boldsymbol{\eta}}(\boldsymbol{L}_i);s)\) and \(\widehat{\gamma}'_l(\boldsymbol{L}_i,d_{\boldsymbol{\eta}}(\boldsymbol{L}_i);s)\) are provided below.

Moreover, \( d\widehat{M}_{C}\{r| Z_i, \boldsymbol{L}_i, d_{\boldsymbol{\eta}}(\boldsymbol{L}_i)\}\) is the martingale increment for the censoring time, namely \( d\widehat{M}_{C}\{r| Z_i, \boldsymbol{L}_i, d_{\boldsymbol{\eta}}(\boldsymbol{L}_i)\}=dN_{C_i}(r)-d\widehat{\Lambda}_C\{r| Z_i, \boldsymbol{L}_i, d_{\boldsymbol{\eta}}(\boldsymbol{L}_i)\}Y_{C_i}(r)\).
\begin{align*}
\widehat{\gamma}_1(\boldsymbol{L}_i,d_{\boldsymbol{\eta}}(\boldsymbol{L}_i);s)=\sum_z&\frac{(2z-1) (2d_{\boldsymbol{\eta}}(\boldsymbol{L}_i)-1) \widehat{\pi}(d_{\boldsymbol{\eta}}(\boldsymbol{L}_i)|{Z_i=z},\boldsymbol{L}_i)}{\widehat{\delta}(\boldsymbol{L}_i)}\\
  &\times \widehat{S}_{T}\left\{s | {Z_i=z}, \boldsymbol{L}_i,d_{\boldsymbol{\eta}}(\boldsymbol{L}_i)\right\}  d \widehat{\Lambda}_{T}\left\{s | {Z_i=z}, \boldsymbol{L}_i,d_{\boldsymbol{\eta}}(\boldsymbol{L}_i)\right\}, 
\end{align*}
\begin{align*}
  \widehat{\gamma}_1^{'}(\boldsymbol{L}_i,d_{\boldsymbol{\eta}}(\boldsymbol{L}_i);s)=&(2d_{\boldsymbol{\eta}}(\boldsymbol{L}_i)-1) \widehat{\pi}(d_{\boldsymbol{\eta}}(\boldsymbol{L}_i)|Z_i=0,\boldsymbol{L}_i)\\
  &\times \widehat{S}_{T}\left\{s | Z_i=0, \boldsymbol{L}_i ,d_{\boldsymbol{\eta}}(\boldsymbol{L}_i)\right\} d \widehat{\Lambda}_{T}\left\{s | Z_i=0, \boldsymbol{L}_i,d_{\boldsymbol{\eta}}(\boldsymbol{L}_i)\right\},
\end{align*}
\begin{align*}
    \widehat{\gamma}_2(\boldsymbol{L}_i,d_{\boldsymbol{\eta}}(\boldsymbol{L}_i);s)=\sum_z&\frac{(2z-1) (2d_{\boldsymbol{\eta}}(\boldsymbol{L}_i)-1) \widehat{\pi}(d_{\boldsymbol{\eta}}(\boldsymbol{L}_i)|{Z_i=z},\boldsymbol{L}_i)}{\widehat{\delta}(\boldsymbol{L}_i)}  \widehat{S}_{T}\left\{s |{Z_i=z}, \boldsymbol{L}_i,d_{\boldsymbol{\eta}}(\boldsymbol{L}_i)\right\}, 
\end{align*}
and
\begin{align*}
  \widehat{\gamma}^{'}_2(\boldsymbol{L}_i,d_{\boldsymbol{\eta}}(\boldsymbol{L}_i);s)=&(2d_{\boldsymbol{\eta}}(\boldsymbol{L}_i)-1) \widehat{\pi}(d_{\boldsymbol{\eta}}(\boldsymbol{L}_i)|Z_i=0,\boldsymbol{L}_i) \widehat{S}_{T}\left\{s | Z_i=0, \boldsymbol{L}_i ,d_{\boldsymbol{\eta}}(\boldsymbol{L}_i)\right\}.
\end{align*}

To incorporate working models like the Cox proportional hazards model, the estimators \(\widehat{E}\{m_l(T_i; s)|T_i\geq r, Z_i, \boldsymbol{L}_i, d_{\boldsymbol{\eta}}(\boldsymbol{L}_i) \},\ 0\leq r<s,\ l=1,2\) can be rewritten as follows.

\begin{align*}
    \widehat{E}\{m_1(T_i;s)|T_i\geq r, Z_i, \boldsymbol{L}_i, d_{\boldsymbol{\eta}}(\boldsymbol{L}_i) \}= d \widehat{\Lambda}_T\{s | Z_i, \boldsymbol{L}_i, d_{\boldsymbol{\eta}}(\boldsymbol{L}_i)\}\frac{\widehat{S}_T\{s | Z_i, \boldsymbol{L}_i, d_{\boldsymbol{\eta}}(\boldsymbol{L}_i)\}}{\widehat{S}_T\{r | Z_i, \boldsymbol{L}_i, d_{\boldsymbol{\eta}}(\boldsymbol{L}_i)\}},
\end{align*}

and 
\begin{align*}
    \widehat{E}\{m_2(T_i;s)|T_i\geq r, Z_i, \boldsymbol{L}_i, d_{\boldsymbol{\eta}}(\boldsymbol{L}_i) \}= \frac{\widehat{S}_T\{s | Z_i, \boldsymbol{L}_i, d_{\boldsymbol{\eta}}(\boldsymbol{L}_i)\}}{\widehat{S}_T\{r | Z_i, \boldsymbol{L}_i, d_{\boldsymbol{\eta}}(\boldsymbol{L}_i)\}}.
\end{align*}

Despite the intricacy of the estimator (\ref{S_D}), we can compute it utilizing estimators \(\widehat{S}_{C}\{s | Z, \boldsymbol{L}, A\}\), \( \widehat{f}(Z | \boldsymbol{L}) \), \( \widehat{\pi}(A|Z,\boldsymbol{L}) \), and \(\widehat{S}_{T}(s | Z, \boldsymbol{L}, A)\). Moreover, it is consistent when either (i) models for \(S_{C}\{s | Z, \boldsymbol{L}, A\}\), \( f(Z | \boldsymbol{L}) \), and \( \pi(A|Z,\boldsymbol{L}) \) or (ii) models for \(S_{T}(s | Z, \boldsymbol{L}, A)\) and \( \pi(A|Z,\boldsymbol{L}) \) are correct. We compare our estimators (\ref{S_I}) and (\ref{S_D}) to other doubly robust estimators in Table \ref{compare}, which indicates that our doubly robust estimator is the only one that has all the desirable properties. 

\begin{table}[!htbp]
 \setlength{\tabcolsep}{3pt}
  \caption{Comparison of various doubly robust estimators and our estimators.}
  \label{compare}
  \centering
  \begin{tabular}{lcccc}
  \toprule & Double  & Robustness to & Robustness to unmeasured  \\
&robustness  & censoring & confounder  \\
  \midrule
    DRKME-IV &\(\checkmark\) & \(\checkmark\)&\(\checkmark\) \\
  IWKME-IV &\(\times\) &\(\checkmark\) &\(\checkmark\) \\
   \cite{jiangEstimationOptimalTreatment2017} & \(\checkmark\) & \(\checkmark\) & \(\times\)  \\
  \cite{cuiSemiparametricInstrumentalVariable2021} &\(\checkmark\) &\(\times\) &\(\checkmark\)\\
  \bottomrule
\end{tabular}
\begin{tablenotes}
  \footnotesize
  \item Note: symbols \(\checkmark\) and \(\times\) denotes yes and no, respectively.
\end{tablenotes}
\end{table}

\subsection{Computational aspect} \label{2.3}
\begin{figure}[!htbp]
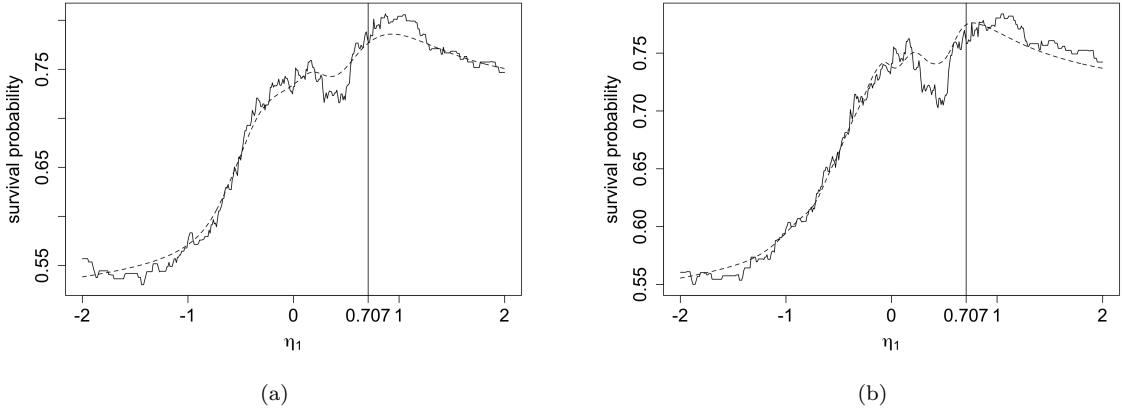

  \begin{center}
  \subfigure[ ]{
  \includegraphics[width=0.45\textwidth]{Fig2a.jpeg}
  }
  \quad
  \subfigure[ ]{
  \includegraphics[width=0.45\textwidth]{Fig2b.jpeg}
  }
  \end{center}
  \caption{Plots for the original (solid line) and smoothed (dashed line) estimates: (a) \(\widehat{S}_{.I}^*(t; \boldsymbol{\eta})\); (b) \(\widehat{S}_{.D}^*(t; \boldsymbol{\eta})\)}
  \label{fig_2}
\end{figure}

The estimators (\ref{S_I})
and (\ref{S_D})
are non-smoothed functions of \( {\boldsymbol{\eta}} .\) To illustrate this point, we plot them as functions of \( {\eta}_{1} \) in Figure \ref{fig_2}, where the optimal value is \(0.707\). As shown in the figure, the estimates are jagged, and direct optimization would lead to a local maximizer. Therefore, we propose to smooth the estimators by using kernel smoothers. Specifically, the indicator function \( d_{\boldsymbol{\eta}}(\boldsymbol{L})=I(\tilde{\boldsymbol{L}}^{T}{\boldsymbol{\eta}}  \geq 0) \) in the IWKME-IV \(\widehat{S}_{I}^*(t ;\boldsymbol{\eta})\) and DRKME-IV \(\widehat{S}_{D}^*(t ;\boldsymbol{\eta})\) is replaced by its smoothed counterpart \( \tilde{d}_{{\boldsymbol{\eta}}}(\boldsymbol{L})=\Phi(\tilde{\boldsymbol{L}}^{T} {\boldsymbol{\eta}}/ h) \) to obtain the smoothed IWKME-IV (SIWKME-IV) \(\widehat{S}_{SI}^*(t ;\boldsymbol{\eta})\) and the SDRKME-IV \(\widehat{S}_{SD}^*(t ;\boldsymbol{\eta})\). The cumulative distribution function \( \Phi(s) \) refers to the standard normal distribution, and the bandwidth parameter \( h \) goes to 0 as \( n \rightarrow \infty \). We set \( h=c_{0} n^{-1 / 3} \operatorname{sd}(\tilde{\boldsymbol{L}}^{T}{\boldsymbol{\eta}}) \), where \( \operatorname{sd}(X) \) is the sample standard deviation of random variable \( X \) and \( c_{0}=4^{1 / 3} \), which performs well in our simulation and application. 
The smoothed estimates are also plotted in Figure \ref{fig_2} to show that smoothed estimates are a great approximation of the original estimates without jaggedness.

Note that a direct replacement of \(d_{\boldsymbol{\eta}}(\boldsymbol{L})\) by \(\Phi(\tilde{\boldsymbol{L}}^{T} {\boldsymbol{\eta}}/ h)\) would cause \(I\{d_{\boldsymbol{\eta}}(\boldsymbol{L})=A\}\) in (\ref{S_I}) and (\ref{S_D}) always being \(0\), making the smoothed estimators questionable. Thus, it is necessary to rewrite \(\widehat{\omega}_i(d_{\boldsymbol{\eta}}(\boldsymbol{L}_i);s)\) in (\ref{S_I}) and \( \widehat{\psi}_{l,i}(d_{\boldsymbol{\eta}}(\boldsymbol{L}_i);s)\) in (\ref{S_D}) as \(\widehat{\omega}_i(d_{\boldsymbol{\eta}}(\boldsymbol{L}_i);s)=d_{\boldsymbol{\eta}}(\boldsymbol{L}_i)\widehat{\omega}_i(A_i=1;s)+(1-d_{\boldsymbol{\eta}}(\boldsymbol{L}_i))\widehat{\omega}_i(A_i=0;s)\) and \( \widehat{\psi}_{l,i}(d_{\boldsymbol{\eta}}(\boldsymbol{L}_i);s)=d_{\boldsymbol{\eta}}(\boldsymbol{L}_i) \widehat{\psi}_{l,i}(A_i=1;s)+(1-d_{\boldsymbol{\eta}}(\boldsymbol{L}_i)) \widehat{\psi}_{l,i}(A_i=0;s)\), respectively. Subsequently, substitute \(d_{\boldsymbol{\eta}}(\boldsymbol{L})\) with \(\Phi(\tilde{\boldsymbol{L}}^{T} {\boldsymbol{\eta}}/ h)\) in the reformulated terms to get \(\widehat{S}_{SI}^*(t;\boldsymbol{\eta})\) and \(\widehat{S}_{SD}^*(t;\boldsymbol{\eta})\).

As the estimates may still be non-concave, we choose to employ the genetic algorithm within the R package \texttt{rgenoud} \cite[]{mebaneGeneticOptimizationUsing2011} to evaluate the performance of our estimators. A genetic algorithm is an optimization method that uses the principles of natural selection and evolution to find approximate solutions to complex problems. The process involves creating a starting population of candidate solutions, evaluating their fitness, selecting the fittest individuals to create the next generation, combining them through crossover, and introducing diversity through mutation. The algorithm repeats this process until a high-fitness solution is found \cite[]{mitchellIntroductionGeneticAlgorithms1998}. The overall algorithm for our two methods is summarized in Algorithm \ref{alg_1}, where, for simplicity, we refer to the first approach as the one maximizing \(\widehat{S}_{SI}^*(t; \boldsymbol{\eta})\) and the second approach as the one maximizing \(\widehat{S}_{SD}^*(t; \boldsymbol{\eta})\).

\begin{algorithm}[!htbp]
  \caption{The algorithm for estimating optimal treatment regimes using an IV.}
  \label{alg_1}
  \hspace*{0.02in} {\bf Input:}
  input \(t\) of interest and the observed data \(\{\boldsymbol{L}_i,A_i,Z_i,\tilde{T}_i,\delta_i\}, i=1, 2, ..., n\).\\
  \hspace*{0.02in} {\bf Output:}
  output the estimated optimal treatment regime \(d_{\widehat{\boldsymbol{\eta}}^{opt}}(\boldsymbol{L})\).
  \begin{algorithmic}[1]
  \State Fit models for nuisance functions to construct estimators \( \widehat{f}(Z | \boldsymbol{L}) \), \( \widehat{\pi}(A|Z,\boldsymbol{L}) \), and \(\widehat{S}_{C}\{s | Z, \boldsymbol{L}, A\}\) for the first approach and an extra estimator \(\widehat{S}_T\{s | Z, \boldsymbol{L}, A\} \) for the second approach, where \(0 \leq s\leq t\) and \(s\in\{\tilde{T}\}_{i=1}^{n}\).
  \vspace{0.2em}
  \State Calculate \(\widehat{\omega}_i(A_i=a;s)\) for the first approach and \(\widehat{\psi}_{l,i}(A_i=a;s)\) for the second approach, where \(i=1,2,...,n\), \(a\in\{0,1\}\), \(l\in\{1,2\}\), \(0 \leq s\leq t\), and \(s\in\{\tilde{T}_i\}_{i=1}^{n}\).
  \vspace{0.2em}
  \State Construct \(\widehat{S}_{SI}^*(t;\boldsymbol{\eta})\) with \(\boldsymbol{\eta}\) as the parameters to be optimized for the first approach and \(\widehat{S}_{SD}^*(t;\boldsymbol{\eta})\) for the second approach.
  \vspace{0.2em}
  \State Use the genetic algorithm to obtain \(\widehat{\boldsymbol{\eta}}^{opt}\) by maximizing \(\widehat{S}_{SI}^*(t ;\boldsymbol{\eta})\) for the first approach and \(\widehat{S}_{SD}^*(t ;\boldsymbol{\eta})\) for the second approach.
  \vspace{0.2em}
  \State Output \(d_{\widehat{\boldsymbol{\eta}}^{opt}}(\boldsymbol{L})=I\{\tilde{\boldsymbol{L}}^{T}\widehat{\boldsymbol{\eta}}^{opt}\geq 0\}\).
  \end{algorithmic}
  \end{algorithm}  

{
\section{Asymptotic properties} \label{sec3}
The following three theorems study the asymptotic properties of our proposed estimators for large samples. To show the asymptotic behavior of the estimator (\ref{S_I}), \(\widehat{S}_I^*(t;\boldsymbol{\eta})\), consider the following conditions. For a function \(g\) of \(\boldsymbol{L}\), define \(\|g(\boldsymbol{L})\|^2=\int\{g(\boldsymbol{l})\}^2dP(\boldsymbol{l})\).

\vspace{1ex}
\textbf{C1:} The estimators \(\widehat{\pi}(A|Z,\boldsymbol{L}), \widehat{f}(Z|\boldsymbol{L}), \widehat{S}_C(s|Z, \boldsymbol{L},A)\) as functions of \(Z,\boldsymbol{L}, A\) are in the Donsker class.

\textbf{C2:} For some constant \( \epsilon>0\), 
\(P(\epsilon<\widehat{\delta}(\boldsymbol{L})<\infty)=1 \), \(P(\epsilon< \widehat{f}(Z=z | \boldsymbol{L})<\infty)=1 \), and \(P(\epsilon< \widehat{S}_C(s | Z=z, \boldsymbol{L}, A=a)<\infty)=1 \) for \(z,a\in\{0,1\}\) and \(0\leq s \leq t\).

\textbf{C3:} \(\| \pi(A=a|Z=z,\boldsymbol{L})-\widehat{\pi}(A=a|Z=z,\boldsymbol{L})\|=o_p(1)\), \(\|f(Z=z | \boldsymbol{L})-\widehat{f}(Z=z | \boldsymbol{L})\|=o_p(1)\), and \(\|S_C(s | Z=z, \boldsymbol{L}, A=a)- \widehat{S}_C(s | Z=z, \boldsymbol{L}, A=a) \|=o_p(1)\) for \(z,a\in\{0,1\}\) and \(0\leq s \leq t\).
\vspace{1ex}

These are standard regularity conditions to establish the convergence result \cite[]{leeDoublyRobustNonparametric2021,kennedySemiparametricTheoryEmpirical2016}. Donsker class (C1) includes regular parametric classes (e.g., the linear model and the generalized linear model) and infinite-dimensional classes (e.g., the single index model and the Cox proportional hazards model) \cite[]{kennedySemiparametricTheoryEmpirical2016}.
Condition C2 is required to ensure that the estimator is well-defined with nonzero denominators. Condition C3 requires that each nuisance model is consistent, which can be guaranteed by a correctly specified model or a nonparametric model.


\begin{theorem}
Under assumptions \(\mathrm{A}1\)-\(\mathrm{A}7\) and conditions \(\mathrm{C}1\)-\(\mathrm{C}3\), the asymptotic property of the estimator (\ref{S_I}), \(\widehat{S}_{I}^{*}\left(t; \boldsymbol{\eta}\right)\), for any given \(\boldsymbol{\eta}\) is as follows:
\begin{align*}   
  \widehat{S}_{I}^{*}\left(t; \boldsymbol{\eta}\right)-S^{*}\left(t; \boldsymbol{\eta}\right)=&O_p\Big\{\sup_{0\leq s \leq t}\sum_z \sum_a\| \delta(\boldsymbol{L}) f(Z=z | \boldsymbol{L}) S_{C}\{s | Z=z, \boldsymbol{L}, 
 A=a\}\\
 &-\widehat{\delta}(\boldsymbol{L}) \widehat{f}(Z=z | \boldsymbol{L}) \widehat{S}_{C}\{s | Z=z, \boldsymbol{L}, A=a\} \|\Big\}+\mathbb{W}(t)+o_p(n^{-1/2}).
\end{align*}
The term \(\mathbb{W}(t)\) given in (1) of the supporting information is asymptotically normal with zero mean.
  \label{the1}
\end{theorem}

Provided that the models for \(\pi(A|Z, \boldsymbol{L})\), \(f(Z|\boldsymbol{L})\), and \(S_{C}\{s | Z=z, \boldsymbol{L}, A=a\}\) converge at a uniform rate \(n^{-1/2}\) such that \(\sup_{0\leq s \leq t}\| \delta(\boldsymbol{L}) f(Z=z | \boldsymbol{L}) S_{C}\{s | Z=z, \boldsymbol{L}, A=a\}-\widehat{\delta}(\boldsymbol{L}) \widehat{f}(Z=z | \boldsymbol{L}) \widehat{S}_{C}\{s | Z=z, \boldsymbol{L}, A=a\}\|=O_p(n^{-1/2})\), the estimator \(\widehat{S}_{I}^{*}\left(t; \boldsymbol{\eta}\right)\) is root-\(n\) consistent. This is often achievable for parametric models, such as the logistic model, and some semiparametric models, such as the Cox proportional hazards model \cite[]{lin2007on}. If we utilize other semiparametric or nonparametric models, such as the single-index model and the random forest, the convergence rate is slower than \(n^{-1/2}\). However, correctly specified semiparametric and nonparametric models can guarantee the consistency of \(\widehat{S}_{I}^{*}\left(t; \boldsymbol{\eta}\right)\). 

To establish the asymptotic and doubly robust properties for the estimator \(\widehat{S}_D^*(t;\boldsymbol{\eta})\), we replace conditions C1-C3 by conditions C1\(^*\)-C3\(^*\) as follows. We also abbreviate \(S_{T,s,z,a}=S_{T}\{s|Z=z,\boldsymbol{L},A=a\}\) and \(S_{C,s,z,a}=S_{C}\{s|Z=z,\boldsymbol{L},A=a\}\) for convenience.

\vspace{1ex}
\textbf{C1\(^*\):} The estimators \(\widehat{\pi}(A|Z,\boldsymbol{L}), \widehat{f}(Z|\boldsymbol{L}), \widehat{S}_C(s|Z, \boldsymbol{L},A), \widehat{S}_T(s|Z, \boldsymbol{L},A)\) as functions of \(Z,\boldsymbol{L}, A\) are in the Donsker class.

\textbf{C2\(^*\):} For some constant \( \epsilon>0\), 
\(P(\epsilon<\widehat{\pi}(A=a|Z=z,\boldsymbol{L})<\infty)=1 \) and \(P(\epsilon< \widehat{f}(Z=z| \boldsymbol{L})<\infty)=1 \) for \(z,a\in\{0,1\}\). Moreover, \(P(\epsilon< \widehat{S}_C(s | Z=z, \boldsymbol{L}, A=a)<\infty)=1 \) and \(P(\epsilon<\widehat{S}_{T}(s|Z=z,\boldsymbol{L},A=a)<\infty)=1 \) for \(z,a \in \{0,1\}\) and \(s\in[0, t] \).

\textbf{C3\(^*\):} \(\| \pi(A=a|Z=z,\boldsymbol{L})-\widehat{\pi}(A=a|Z=z,\boldsymbol{L})\|=o_p(1)\), \(\|f(Z=z | \boldsymbol{L})-\widehat{f}(Z=z | \boldsymbol{L})\|=o_p(1)\), \(\|S_{C,s,z,a}-\widehat{S}_{C,s,z,a}\|=o_p(1)\) or \(\| \pi(A=a|Z=z,\boldsymbol{L})-\widehat{\pi}(A=a|Z=z,\boldsymbol{L})\|=o_p(1)\), \(\|S_{T,s,z,a}-\widehat{S}_{T,s,z,a}\|=o_p(1)\) for \(z,a\in\{0,1\}\) and \(0\leq s \leq t\).
\vspace{1ex}

Compared to conditions C1 and C2, conditions C1\(^*\) and C2\(^*\) additionally restrict the nuisance model \(\widehat{S}_{T,s,z,a}\) in the Donsker class and limit its range. Condition C3\(^*\) is implied by and therefore more general than condition C3.

\begin{theorem} \label{the2}
  Under assumptions \(\mathrm{A}1\)-\(\mathrm{A}7\) and conditions \(\mathrm{C}1^*\)-\(\mathrm{C}3^*\), the asymptotic property of the estimator (\ref{S_D}), \(\widehat{S}_D^*(t;\boldsymbol{\eta})\), for any given \(\boldsymbol{\eta}\) is as follows:
  \begin{align*}
    &\widehat{S}_D^*(t;\boldsymbol{\eta})-S^*(t;\boldsymbol{\eta})\\
  =&O_p\Big\{ \sup_{0\leq s \leq t} \sum_{z} \sum_{a} \|\pi(A=a|Z=z,\boldsymbol{L})S_{T,s,z,a}-\widehat{\pi}(A=a|Z=z,\boldsymbol{L})\widehat{S}_{T,s,z,a}\|\\
  \cdot& \|\delta(\boldsymbol{L})f(Z=z|\boldsymbol{L})-\widehat{\delta}(\boldsymbol{L})\widehat{f}(Z=z|\boldsymbol{L})\| + \|S_{T,s,z,a}-\widehat{S}_{T,s,z,a}\|\|S_{C,s,z,a}-\widehat{S}_{C,s,z,a}\|\Big\}\\
  &+\mathbb{W}(t)+o_p(n^{-1/2}).
\end{align*}
\end{theorem}

Theorem \ref{the2} implies double robustness when either (i) models for \(S_{C}\{s | Z, \boldsymbol{L}, A\}\), \( f(Z | \boldsymbol{L}) \), and \( \pi(A|Z,\boldsymbol{L}) \) or (ii) models for \(S_{T}(s | Z, \boldsymbol{L}, A)\) and \( \pi(A|Z,\boldsymbol{L}) \) are correct. 
It also suggests that the \(n^{-1/4}\) convergence rate for the models \(S_{T}(s | Z, \boldsymbol{L}, A)\), \(S_{C}\{s | Z, \boldsymbol{L}, A\}\), \( \pi(A|Z,\boldsymbol{L}) \) and \(f(Z|\boldsymbol{L})\) is sufficient to guarantee the \(n^{-1/2}\) convergence rate of \(\widehat{S}_D^*(t;\boldsymbol{\eta})\). 
The convergence rate of \(\widehat{S}_D^*(t;\boldsymbol{\eta})\) is therefore no worse than \(\widehat{S}_I^*(t;\boldsymbol{\eta})\), and semiparametric or nonparametric models applied to estimate \(S_{T}(s | Z, \boldsymbol{L}, A)\), \(S_C(s | Z, \boldsymbol{L}, A)\), \( \pi(A|Z,\boldsymbol{L}) \), and \(f(Z|\boldsymbol{L})\) are possible to achieve the \(n^{-1/2}\) convergence rate of \(\widehat{S}_D^*(t;\boldsymbol{\eta})\). For instance, the single index model converging at a rate of \(n^{-2/5}\) \cite[]{horowitz2009semiparametric} and the random forest converging at a rate faster than \(n^{-1/4}\) in very low dimensional settings \cite[]{cui2017tree} can be utilized.

The following theorem asserts that the smoothed estimators possess similar properties to their nonsmoothed counterparts.
\begin{theorem} \label{the3}
  Under the assumptions and conditions of Theorem \ref*{the1} and Theorem \ref*{the2} plus additional conditions in the supporting information, the asymptotic properties of the smoothed estimators are as follows:

  \((1):\) \(\sup_{{\|{\boldsymbol{\eta}}\|_2}=1}\sqrt{n}(\widehat{S}_{I}^*(t ;\boldsymbol{\eta})-\widehat{S}_{SI}^*(t ;\boldsymbol{\eta}))=o_p(1)\);

  \((2):\) \(\sup_{{\|{\boldsymbol{\eta}}\|_2}=1}\sqrt{n}(\widehat{S}_{D}^*(t ;\boldsymbol{\eta})-\widehat{S}_{SD}^*(t ;\boldsymbol{\eta}))=o_p(1)\).
\end{theorem}

}
\section{Numerical studies} \label{4}
We present three simulation studies. In the first study, we analyze the finite-sample performance of our method under different scenarios. In the second study, we examine the performance of our method with different distributions of the unmeasured confounder. In the last study, we vary the censoring mechanisms to validate our method under covariate-dependent censoring. 
\subsection{Monte Carlo simulation} \label{4.1}
The IV \(Z\) is generated from a Bernoulli distribution with probability \(1/2\) to simulate a randomized trial with non-adherence. The simulation under the covariate-dependent IV is placed in the supporting information (F.1). The baseline variable \( \boldsymbol{L} \) is drawn from a uniform distribution on \( [-2,2]^{2} \); \( U \) is obtained from a bridge distribution with the parameter \( \phi=1 / 2 \); treatment \( A \) is generated from a logistic regression with success probability
\[
  \pi(A=1 | Z,\boldsymbol{L}, U)=\operatorname{expit}\left\{-2.5+L_{1}+5 Z-0.5 U\right\}.
\]

The generation mechanism of \(U\) ensures that there exists a vector \( \boldsymbol{\alpha} \) such that \( \operatorname{logit}\{\pi(A= 1 | Z,\boldsymbol{L})\}=(1, Z, \boldsymbol{L}^{T})\boldsymbol{\alpha} \) \cite[]{wangMatchingConditionalMarginal2003}. While a logistic model is correctly specified to estimate \(\pi(A | \boldsymbol{L}, Z)\), note that the logistic model used to generate \(A\) does not satisfy assumption A7. Nevertheless, we still employ the logistic model because it is common in reality. Moreover, the simulation study can serve as an investigation of the consequences of violating assumption A7. We also provide a data generation process satisfying assumption A7 in the supporting information (F.2), which exhibits good performance as expected.
The survival time \( T \) is generated from a linear transformation model \cite[]{jiangEstimationOptimalTreatment2017}, in which the coefficient of \(U\) is varied to simulate the magnitude of the unmeasured confounding. 

\vspace{1ex}
(a) \( h(T)=-0.5 L_{1}+A\left(L_{1}-L_{2}\right)+0.5U+\varepsilon_1 \),

(b) \( h(T)=-0.5 L_{1}+A\left(L_{1}-L_{2}\right)+U+\varepsilon_1 \),

(c) \( h(T)=-0.5 L_{1}+A\left(L_{1}-L_{2}\right)+0.5U+\varepsilon_2 \),

(d) \( h(T)=-0.5 L_{1}+A\left(L_{1}-L_{2}\right)+U+\varepsilon_2 \),
\vspace{1ex}

\noindent where \( h(s)=\log \{\exp (s)-1\}-2 \) is an increasing function. The error terms \( \varepsilon _1\) and \( \varepsilon _2\) follow the extreme value distribution and the logistic distribution, which correspond to the Cox proportional hazards model and the proportional odds model, separately. The covariate-independent censoring time \( C \) is uniformly distributed over \( \left[0, C_{0}\right] \), with \( C_{0} \) chosen to achieve a censoring rate of \( 15 \% \) or \( 30 \% \). Additional censoring settings, the covariate-dependent censoring, can be found in Section \ref{4.3}. We estimate the optimal treatment regime within the class of linear regimes \( \mathcal{D}=\{d_{\boldsymbol{\eta}}: d_{\boldsymbol{\eta}}(L_{1}, L_{2})=I\{{\eta}_{0}+{\eta}_{1} L_{1}+{\eta}_{2} L_{2} \geq 0\}, {\boldsymbol{\eta}}=({\eta}_{0}, {\eta}_{1}, {\eta}_{2})^{\mathrm{T}},\|{\boldsymbol{\eta}}\|_2=1\} \) by maximizing the estimates of the counterfactual survival function \(S^*(t;\boldsymbol{\eta})\) at \(t=2\). The class of linear regimes contains the true optimal treatment regime \( d_{\boldsymbol{\eta}^{opt}}(L_{1}, L_{2})=I\{0.707L_{1}-0.707L_{2} \geq 0\} \).

The proposed estimators are implemented with \( \widehat{\delta}(\boldsymbol{L})=\widehat{\pi}(A=1 |  Z=1,\boldsymbol{L})-\widehat{\pi}(A=1 | Z=0,\boldsymbol{L}) \) and \(\widehat{f}(Z | \boldsymbol{L})\), where \(\widehat{\pi}(A |Z, \boldsymbol{L})\) and \(\widehat{f}(Z | \boldsymbol{L})\) are estimated from logistic models with covariates \((Z, \boldsymbol{L}^T)\) and \((\boldsymbol{L}^T)\), respectively. The bridge distribution of \(U\) ensures that the logistic model for \(\widehat{\pi}(A | \boldsymbol{L}, Z)\) is correctly specified. The Cox proportional hazards model with covariates \((Z, \boldsymbol{L}^{T}, A)\) and \((Z, \boldsymbol{L}^{T}, A, \boldsymbol{L}^TA)\) is posited to estimate \(S_{C}\{s | Z, \boldsymbol{L}, A\}\) and \(S_{T}\{s | Z, \boldsymbol{L}, A\}\), separately. 
Due to the presence of the unmeasured confounder, \(S_{T}\{s | Z, \boldsymbol{L}, A\}\) is misspecified in all settings. Thus, the simulation results will test the double robustness of \(\widehat{S}_{SD}^*(t,\boldsymbol{\eta})\). The simulation is conducted with a sample size of 500, which is repeated 500 times, and the performance is evaluated using a large independent test set comprising of 10000 subjects. Additional simulation results with sample sizes of 250 and 1000 are shown in the supporting information (F.3).

In this article, we use the following three performance indicators: the bias of the estimated \(\widehat{\boldsymbol{\eta}}^{opt}_.\), the mean of the \(t\)-year survival probability following the estimated optimal treatment regime \( S^*(t;{\widehat{\boldsymbol{\eta}}}^{opt}_.) \), and the mean of the misclassification rate by comparing the true and estimated optimal treatment regimes (denoted by MR). The numbers given in parentheses are the standard deviations of the corresponding estimates. Here, \( S^*(t;{\widehat{\boldsymbol{\eta}}}^{opt}_.) \) 
are computed by using simulated survival times following the given treatment regime based on the test set. The MR is calculated as the percentage of patients whose true and estimated optimal treatment regimes differ on the test set. The methods for estimating optimal treatment regimes under comparison are all value search methods which maximize an estimator of the value function within a class of linear regimes. The estimators for the value function include:

\vspace{1ex}
(1) Our proposed estimator, SIWKME-IV, defined by (\ref{S_I}) with a kernel smoother in Section \ref{2.3};

(2) The doubly robust version of the estimator in (1), SDRKME-IV, defined by (\ref{S_D}) with a kernel smoother in Section \ref{2.3};

(3) The smoothed inverse propensity score weighted Kaplan-Meier estimator without adjustment for unmeasured confounders (denoted as SIWKME) \cite[]{jiangEstimationOptimalTreatment2017};

(4) The smoothed augmented inverse propensity score weighted Kaplan-Meier estimator without adjustment for unmeasured confounders (denoted as SAIWKME) \cite[]{jiangEstimationOptimalTreatment2017}, which is the doubly robust version of the SIWKME.
\vspace{1ex}

As shown in Table \ref{tab1}, our method demonstrates more accurate parameter estimates, higher \(t\)-year survival probability, and lower MRs compared to the SIWKME and the SAIWKME. Although the posited model for \(S_{T}\{s | Z, \boldsymbol{L}, A\}\) is misspecified (since we model it without the unmeasured confounder \(U\)), the SDRKME-IV remains consistent as expected and shows superior performance. The excellent performance suggests that even a misspecified model for \(S_{T}\{s | Z, \boldsymbol{L}, A\}\) can still result in improved performance for the SDRKME-IV. The standard deviation of the SDRKME-IV is smaller than that of the SIWKME-IV, which is another advantage of the SDRKME-IV. As illustrated in the supporting information (F.4), a more dominant IV can further reduce the standard deviation.
In settings (b) and (d), where the unmeasured confounding is stronger than that in settings (a) and (c), the SIWKME and SAIWKME deteriorate while our method remains stable. 
On the aspect of censoring rates, lower censoring rates always provide better performance, while the impact of the censoring rates on performance is limited.

\begin{table}[!htbp] 
 \setlength{\tabcolsep}{3pt}
    \caption{Bias (with standard deviation in the parenthesis) of \(\widehat{\boldsymbol{\eta}}_.^{opt}\) relative to \({\boldsymbol{\eta}}_.^{opt}\) and mean (with standard deviation in the parenthesis) of value functions and missing classification rates for Section \ref{4.1}. C\% represents the censoring rate, and MR denotes the misclassification rate. The methods denoted by bold symbols represent our proposed method and the others refer to the methods that do not consider any unmeasured confounding effects. }
    \centering
    \begin{tabular}{lrrrrrrr}
    \toprule Setting &  Method  & C\% & Bias in \( {\boldsymbol{\eta}}_0^{opt} \) & Bias in \( {\boldsymbol{\eta}}_{1}^{opt} \) &Bias in \({\boldsymbol{\eta}}_{2}^{opt}\)& \( S^*(t;\widehat{\boldsymbol{\eta}}_.^{opt})^\dagger  \)& MR \\
    \midrule \multirow{8}{*}{(a)}& \textbf{SIWKME-IV}& $30$ & $ 0.020$ ($ 0.349$) & $-0.068$ ($ 0.217$) & $ 0.101$ ($ 0.235$) & $ 0.736$ ($ 0.013$) &$ 0.142$ ($ 0.073$)\\
  &\textbf{SDRKME-IV}& $30 $ & $-0.007$ ($ 0.332$) & $-0.068$ ($ 0.206$) & $ 0.084$ ($ 0.225$) & $\boldsymbol{ 0.738}$ ($ 0.013$) &$ \boldsymbol{0.134}$ ($ 0.073$)\\
  &SIWKME& $30 $ & $-0.496$ ($ 0.196$) & $-0.132$ ($ 0.159$) & $ 0.134$ ($ 0.178$) & $ 0.728$ ($ 0.015$) &$ 0.199$ ($ 0.078$)\\
  &SAIWKME& $30 $ & $-0.501$ ($ 0.187$) & $-0.132$ ($ 0.155$) & $ 0.135$ ($ 0.179$) & $ 0.728$ ($ 0.015$) &$ 0.200$ ($ 0.076$)\\
\cmidrule{2-8} & \textbf{SIWKME-IV}& $ 15 $ & $ 0.004$ ($ 0.341$) & $-0.055$ ($ 0.203$) & $ 0.099$ ($ 0.222$) & $ 0.737$ ($ 0.013$) &$ 0.137$ ($ 0.071$)\\
 &\textbf{SDRKME-IV}& $15 $ & $-0.012$ ($ 0.335$) & $-0.073$ ($ 0.202$) & $ 0.073$ ($ 0.207$) & $\boldsymbol{ 0.738}$ ($ 0.012$) &$ \boldsymbol{0.133}$ ($ 0.069$)\\
 &SIWKME& $15 $ & $-0.491$ ($ 0.193$) & $-0.131$ ($ 0.152$) & $ 0.126$ ($ 0.175$) & $ 0.729$ ($ 0.015$) &$ 0.197$ ($ 0.076$)\\
 &SAIWKME& $15 $ & $-0.494$ ($ 0.187$) & $-0.133$ ($ 0.155$) & $ 0.125$ ($ 0.169$) & $ 0.729$ ($ 0.014$) &$ 0.197$ ($ 0.074$)\\
\midrule\multirow{8}{*}{(b)} & \textbf{SIWKME-IV}& $30 $ & $ 0.026$ ($ 0.418$) & $-0.116$ ($ 0.291$) & $ 0.149$ ($ 0.282$) & $ \boldsymbol{0.657}$ ($ 0.015$) &$ 0.178$ ($ 0.095$)\\
 &\textbf{SDRKME-IV}& $30 $ & $ 0.002$ ($ 0.414$) & $-0.119$ ($ 0.294$) & $ 0.143$ ($ 0.281$) & $ \boldsymbol{0.657}$ ($ 0.014$) &$ \boldsymbol{0.177}$ ($ 0.096$)\\
 &SIWKME& $30 $ & $-0.701$ ($ 0.194$) & $-0.299$ ($ 0.228$) & $ 0.254$ ($ 0.218$) & $ 0.635$ ($ 0.022$) &$ 0.317$ ($ 0.108$)\\
 &SAIWKME& $30 $ & $-0.709$ ($ 0.180$) & $-0.297$ ($ 0.219$) & $ 0.256$ ($ 0.214$) & $ 0.635$ ($ 0.022$) &$ 0.320$ ($ 0.106$)\\
\cmidrule{2-8} & \textbf{SIWKME-IV}& $15 $ & $ 0.018$ ($ 0.422$) & $-0.112$ ($ 0.281$) & $ 0.152$ ($ 0.284$) & $ 0.657$ ($ 0.014$) &$ 0.179$ ($ 0.092$)\\
 &\textbf{SDRKME-IV}& $15 $ & $-0.008$ ($ 0.410$) & $-0.107$ ($ 0.278$) & $ 0.135$ ($ 0.260$) & $ \boldsymbol{0.658}$ ($ 0.013$) &$ \boldsymbol{0.172}$ ($ 0.088$)\\
 &SIWKME& $15 $ & $-0.705$ ($ 0.188$) & $-0.290$ ($ 0.217$) & $ 0.259$ ($ 0.216$) & $ 0.635$ ($ 0.021$) &$ 0.318$ ($ 0.107$)\\
 &SAIWKME& $15 $ & $-0.711$ ($ 0.176$) & $-0.293$ ($ 0.214$) & $ 0.257$ ($ 0.209$) & $ 0.635$ ($ 0.021$) &$ 0.319$ ($ 0.105$)\\
\midrule\multirow{8}{*}{(c)} & \textbf{SIWKME-IV}& $30 $ & $ 0.023$ ($ 0.393$) & $-0.095$ ($ 0.256$) & $ 0.123$ ($ 0.255$) & $ 0.619$ ($ 0.016$) &$ 0.161$ ($ 0.087$)\\
 &\textbf{SDRKME-IV}& $30 $ & $-0.007$ ($ 0.372$) & $-0.093$ ($ 0.261$) & $ 0.110$ ($ 0.246$) & $ \boldsymbol{0.620}$ ($ 0.014$) &$ \boldsymbol{0.157}$ ($ 0.081$)\\
 &SIWKME& $30 $ & $-0.484$ ($ 0.231$) & $-0.151$ ($ 0.192$) & $ 0.136$ ($ 0.200$) & $ 0.613$ ($ 0.015$) &$ 0.205$ ($ 0.082$)\\
 &SAIWKME& $30 $ & $-0.489$ ($ 0.225$) & $-0.149$ ($ 0.190$) & $ 0.139$ ($ 0.203$) & $ 0.613$ ($ 0.015$) &$ 0.205$ ($ 0.081$)\\
\cmidrule{2-8} & \textbf{SIWKME-IV}& $15 $ & $ 0.031$ ($ 0.380$) & $-0.098$ ($ 0.259$) & $ 0.107$ ($ 0.239$) & $ 0.620$ ($ 0.015$) &$ 0.158$ ($ 0.081$)\\
 &\textbf{SDRKME-IV}& $15 $ & $ 0.004$ ($ 0.366$) & $-0.101$ ($ 0.256$) & $ 0.090$ ($ 0.232$) & $ \boldsymbol{0.621}$ ($ 0.013$) &$ \boldsymbol{0.154}$ ($ 0.076$)\\
 &SIWKME& $15 $ & $-0.482$ ($ 0.229$) & $-0.144$ ($ 0.184$) & $ 0.135$ ($ 0.194$) & $ 0.614$ ($ 0.015$) &$ 0.201$ ($ 0.081$)\\
 &SAIWKME& $15 $ & $-0.483$ ($ 0.228$) & $-0.145$ ($ 0.182$) & $ 0.134$ ($ 0.193$) & $ 0.614$ ($ 0.015$) &$ 0.201$ ($ 0.082$)\\
\midrule\multirow{8}{*}{(d)} & \textbf{SIWKME-IV}& $30 $ & $ 0.049$ ($ 0.449$) & $-0.151$ ($ 0.317$) & $ 0.167$ ($ 0.309$) & $ \boldsymbol{0.582}$ ($ 0.016$) &$ 0.198$ ($ 0.102$)\\
&\textbf{SDRKME-IV}& $30 $ & $ 0.036$ ($ 0.443$) & $-0.142$ ($ 0.312$) & $ 0.172$ ($ 0.315$) & $ \boldsymbol{0.582}$ ($ 0.016$) &$ \boldsymbol{0.195}$ ($ 0.102$)\\
&SIWKME& $30 $ & $-0.679$ ($ 0.202$) & $-0.289$ ($ 0.231$) & $ 0.242$ ($ 0.231$) & $ 0.567$ ($ 0.020$) &$ 0.306$ ($ 0.108$)\\
&SAIWKME& $30 $ & $-0.685$ ($ 0.197$) & $-0.290$ ($ 0.229$) & $ 0.245$ ($ 0.229$) & $ 0.567$ ($ 0.020$) &$ 0.308$ ($ 0.108$)\\
\cmidrule{2-8} & \textbf{SIWKME-IV}& $15 $ & $ 0.042$ ($ 0.436$) & $-0.129$ ($ 0.298$) & $ 0.175$ ($ 0.320$) & $ \boldsymbol{0.583}$ ($ 0.016$) &$ 0.193$ ($ 0.101$)\\
 &\textbf{SDRKME-IV}& $15 $ & $ 0.036$ ($ 0.432$) & $-0.135$ ($ 0.301$) & $ 0.163$ ($ 0.314$) & $ \boldsymbol{0.583}$ ($ 0.015$) &$ \boldsymbol{0.192}$ ($ 0.099$)\\
 &SIWKME& $15 $ & $-0.691$ ($ 0.191$) & $-0.291$ ($ 0.233$) & $ 0.253$ ($ 0.228$) & $ 0.567$ ($ 0.020$) &$ 0.311$ ($ 0.107$)\\
 &SAIWKME& $15 $ & $-0.695$ ($ 0.187$) & $-0.289$ ($ 0.228$) & $ 0.253$ ($ 0.222$) & $ 0.567$ ($ 0.020$) &$ 0.312$ ($ 0.106$)\\
\bottomrule
\end{tabular}
\begin{tablenotes}
  \footnotesize
  \item \(\dagger \)\( S^*(t;\boldsymbol{\eta}^{opt}) \) are \(0.750, 0.671, 0.634, 0.598\) for four settings, respectively.
\end{tablenotes}
\label{tab1}
\end{table}

\subsection{Simulation under different distributions of the unmeasured confounder} \label{4.2}
In the second numerical study, we demonstrate the performance of our method under different distributions of the unmeasured confounder. Specifically, \(U\) is generated from a bridge distribution with the parameter \(\phi=1/2\), a normal distribution \(N(0,3.14^2)\), and a uniform distribution on \([-5.44,5.44]\), respectively, all of which have the same standard deviation \(3.14\). Note that, unlike the bridge distribution, applying the logistic model to estimate \(\pi(A | \boldsymbol{L}, Z)\) is misspecified for other distributions. Thus, the results can also be considered as a simulation study to evaluate the impact of a misspecified model for \(\pi(A | \boldsymbol{L}, Z)\). The remaining generation is the same as the settings (a) and (c) in Section \ref{4.1} with a fixed censoring rate of 15\%. We also maintain the same estimation procedure as that in Section \ref{4.1}. In view of the result in Table \ref{tab2}, the violation of assumption A7 and an incorrectly posited model for \(\pi(A | \boldsymbol{L}, Z)\) have a limited impact.


\begin{table}[!htbp]
 \setlength{\tabcolsep}{3pt}
  \caption{Bias (with standard deviation in the parenthesis) of \(\widehat{\boldsymbol{\eta}}_.^{opt}\) relative to \({\boldsymbol{\eta}}_.^{opt}\) and mean (with standard deviation in the parenthesis) of value functions and missing classification rates for Section \ref{4.2}. MR denotes the misclassification rate. The methods denoted by bold symbols represent our proposed method and the others refer to the methods that do not consider any unmeasured confounding effects.
  }
  \centering
  \begin{tabular}{lrrrrrrr}
  \toprule Case &  Method  & Distribution\(^\ast \) & Bias in \( {\boldsymbol{\eta}}_0^{opt} \) & Bias in \( {\boldsymbol{\eta}}_{1}^{opt} \) &Bias in \({\boldsymbol{\eta}}_{2}^{opt}\)& \( S^*(t;\widehat{\boldsymbol{\eta}}_.^{opt})^\dagger\)&MR \\
  \cmidrule{1-8} \multirow{12}{*}{(a)} & \textbf{SIWKME-IV}&  bridge  & $ 0.004$ ($ 0.341$) & $-0.055$ ($ 0.203$) & $ 0.099$ ($ 0.222$) & $ 0.737$ ($ 0.013$) &$ 0.137$ ($ 0.071$)\\
 &\textbf{SDRKME-IV}& bridge & $-0.012$ ($ 0.335$) & $-0.073$ ($ 0.202$) & $ 0.073$ ($ 0.207$) & $\boldsymbol{ 0.738}$ ($ 0.012$) &$ \boldsymbol{0.133}$ ($ 0.069$)\\
 &SIWKME& bridge & $-0.491$ ($ 0.193$) & $-0.131$ ($ 0.152$) & $ 0.126$ ($ 0.175$) & $ 0.729$ ($ 0.015$) &$ 0.197$ ($ 0.076$)\\
 &SAIWKME& bridge & $-0.494$ ($ 0.187$) & $-0.133$ ($ 0.155$) & $ 0.125$ ($ 0.169$) & $ 0.729$ ($ 0.014$) &$ 0.197$ ($ 0.074$)\\
  \cmidrule{2-8}& \textbf{SIWKME-IV}& normal & $-0.007$ ($ 0.361$) & $-0.077$ ($ 0.226$) & $ 0.095$ ($ 0.218$) & $ 0.719$ ($ 0.012$) &$ 0.146$ ($ 0.071$)\\
  &\textbf{SDRKME-IV}& normal & $-0.014$ ($ 0.353$) & $-0.083$ ($ 0.225$) & $ 0.084$ ($ 0.218$) & $ \boldsymbol{0.720}$ ($ 0.011$) &$ \boldsymbol{0.144}$ ($ 0.069$)\\
  &SIWKME& normal & $-0.546$ ($ 0.197$) & $-0.170$ ($ 0.170$) & $ 0.146$ ($ 0.178$) & $ 0.707$ ($ 0.018$) &$ 0.224$ ($ 0.086$)\\
  &SAIWKME& normal & $-0.547$ ($ 0.196$) & $-0.173$ ($ 0.169$) & $ 0.140$ ($ 0.168$) & $ 0.707$ ($ 0.017$) &$ 0.223$ ($ 0.085$)\\
  \cmidrule{2-8}&\textbf{SIWKME-IV}& uniform & $ 0.022$ ($ 0.383$) & $-0.093$ ($ 0.256$) & $ 0.113$ ($ 0.240$) & $ 0.705$ ($ 0.014$) &$ 0.156$ ($ 0.082$)\\
  &\textbf{SDRKME-IV}& uniform & $-0.002$ ($ 0.364$) & $-0.088$ ($ 0.249$) & $ 0.099$ ($ 0.232$) & $ \boldsymbol{0.706}$ ($ 0.013$) &$\boldsymbol{ 0.150}$ ($ 0.079$)\\
  &SIWKME& uniform & $-0.615$ ($ 0.178$) & $-0.231$ ($ 0.176$) & $ 0.157$ ($ 0.175$) & $ 0.690$ ($ 0.018$) &$ 0.255$ ($ 0.086$)\\
  &SAIWKME& uniform & $-0.615$ ($ 0.170$) & $-0.227$ ($ 0.167$) & $ 0.153$ ($ 0.170$) & $ 0.690$ ($ 0.018$) &$ 0.253$ ($ 0.084$)\\
  \midrule \multirow{12}{*}{(c)}& \textbf{SIWKME-IV}& bridge & $ 0.031$ ($ 0.380$) & $-0.098$ ($ 0.259$) & $ 0.107$ ($ 0.239$) & $ 0.620$ ($ 0.015$) &$ 0.158$ ($ 0.081$)\\
 &\textbf{SDRKME-IV}& bridge & $ 0.004$ ($ 0.366$) & $-0.101$ ($ 0.256$) & $ 0.090$ ($ 0.232$) & $ \boldsymbol{0.621}$ ($ 0.013$) &$ \boldsymbol{0.154}$ ($ 0.076$)\\
  &SIWKME& bridge & $-0.484$ ($ 0.231$) & $-0.151$ ($ 0.192$) & $ 0.136$ ($ 0.200$) & $ 0.613$ ($ 0.015$) &$ 0.205$ ($ 0.082$)\\
 &SAIWKME& bridge & $-0.483$ ($ 0.228$) & $-0.145$ ($ 0.182$) & $ 0.134$ ($ 0.193$) & $ 0.614$ ($ 0.015$) &$ 0.201$ ($ 0.082$)\\
  \cmidrule{2-8}&\textbf{SIWKME-IV}& normal & $ 0.021$ ($ 0.391$) & $-0.085$ ($ 0.248$) & $ 0.128$ ($ 0.250$) & $ \boldsymbol{0.612}$ ($ 0.014$) &$ 0.162$ ($ 0.081$)\\
  &\textbf{SDRKME-IV}& normal & $ 0.008$ ($ 0.389$) & $-0.087$ ($ 0.243$) & $ 0.119$ ($ 0.244$) & $ \boldsymbol{0.612}$ ($ 0.013$) &$ \boldsymbol{0.159}$ ($ 0.080$)\\
  &SIWKME&normal & $-0.535$ ($ 0.214$) & $-0.169$ ($ 0.192$) & $ 0.152$ ($ 0.187$) & $ 0.602$ ($ 0.017$) &$ 0.222$ ($ 0.090$)\\
  &SAIWKME& normal & $-0.530$ ($ 0.218$) & $-0.171$ ($ 0.194$) & $ 0.149$ ($ 0.188$) & $ 0.602$ ($ 0.017$) &$ 0.222$ ($ 0.090$)\\
  \cmidrule{2-8}&\textbf{SIWKME-IV}& uniform  & $ 0.034$ ($ 0.417$) & $-0.123$ ($ 0.281$) & $ 0.133$ ($ 0.277$) & $ \boldsymbol{0.609}$ ($ 0.017$) &$ 0.175$ ($ 0.093$)\\
  &\textbf{SDRKME-IV}& uniform & $ 0.007$ ($ 0.399$) & $-0.119$ ($ 0.284$) & $ 0.123$ ($ 0.273$) & $ \boldsymbol{0.609}$ ($ 0.016$) &$ \boldsymbol{0.171}$ ($ 0.088$)\\
  &SIWKME& uniform & $-0.589$ ($ 0.203$) & $-0.202$ ($ 0.188$) & $ 0.176$ ($ 0.198$) & $ 0.599$ ($ 0.017$) &$ 0.247$ ($ 0.089$)\\
  &SAIWKME& uniform & $-0.590$ ($ 0.195$) & $-0.201$ ($ 0.186$) & $ 0.172$ ($ 0.194$) & $ 0.599$ ($ 0.017$) &$ 0.246$ ($ 0.088$)\\
  \bottomrule
\end{tabular}
\begin{tablenotes}
  \footnotesize
  \item \(\ast \) Distribution, the distribution for generating \(U\) is categorized into three variants: obtained from the bridge distribution, the normal distribution, and the uniform distribution.
  \item \(\dagger \) The values of \( S^*(t;\boldsymbol{\eta}^{opt}) \) for the two settings with the bridge distribution are 0.750 and 0.634, whereas for the normal distribution, they are 0.733 and 0.626, and for the uniform distribution, they are 0.719 and 0.625, respectively.
\end{tablenotes}
\label{tab2}
\end{table}

\subsection{Simulation with covariate-dependent censoring} \label{4.3}
In this section, we use the same data generation as in settings (a) and (c) of Section \ref{4.1}, but with different censoring schemes: (i) \(h(C)=2+L_1+L_2 + A + \varepsilon_1\); (ii) \(h(C)=2+L_1+L_2 + A + \varepsilon_2\), where \( h(s)=\log \{\exp (s)-1\}-2 \). The error terms \( \varepsilon _1\) and \( \varepsilon _2\) follow separately the extreme value distribution and the logistic distribution, corresponding to a Cox proportional hazards model and a proportional odds model, respectively. The censoring rates for schemes (i) and (ii) in setting (a) are \(0.255\) and \(0.322\), respectively. In setting (b), they are \(0.382\), and \(0.426\), respectively. We maintain the same estimation procedure as in Section \ref{4.1}, where the survival function of the censoring time \(S_C(s|Z,\boldsymbol{L},A)\) is estimated by positing the Cox proportional hazards model. Thus, scheme (i) corresponds to a correctly posited model for \(S_C(s|Z,\boldsymbol{L},A)\), while scheme (ii) corresponds to an incorrectly posited model. As shown in Table \ref{tab7}, our methodology shows satisfactory performance in all scenarios, and the doubly robust estimator always provides better performance. In scheme (ii), although the Cox proportional hazards model is not correctly specified to estimate the survival function of the censoring time, the SIWKME-IV still demonstrates reasonable performance due to its strong predictive ability.

\begin{table}[!htbp]
 \setlength{\tabcolsep}{3pt}
  \caption{Bias (with standard deviation in the parenthesis) of \(\widehat{\boldsymbol{\eta}}_.^{opt}\) relative to \({\boldsymbol{\eta}}_.^{opt}\) and mean (with standard deviation in the parenthesis) of value functions and missing classification rates for Section \ref{4.3}. Censoring represents the different censoring schemes, and MR denotes the misclassification rate. The methods denoted by bold symbols represent our proposed method and the others refer to the methods that do not consider any unmeasured confounding effects.
  }
    \centering
    \begin{tabular}{lrrrrrrr}
    \toprule Case &  Method  & Censoring & Bias in \( {\boldsymbol{\eta}}_0^{opt} \) & Bias in \( {\boldsymbol{\eta}}_{1}^{opt} \) &Bias in \({\boldsymbol{\eta}}_{2}^{opt}\)&\( S^*(t;\widehat{\boldsymbol{\eta}}_.^{opt})^\dagger \)&MR \\
    \midrule \multirow{8}{*}{(a)} & \textbf{SIWKME-IV}& (i) & $-0.014$ ($ 0.333$) & $-0.052$ ($ 0.200$) & $ 0.096$ ($ 0.214$) & $ 0.738$ ($ 0.011$) &$ 0.134$ ($ 0.067$)\\
    &\textbf{SDRKME-IV}&  (i) & $-0.034$ ($ 0.322$) & $-0.047$ ($ 0.191$) & $ 0.091$ ($ 0.211$) & $ \boldsymbol{0.739}$ ($ 0.011$) &$ \boldsymbol{0.131}$ ($ 0.067$)\\
    &SIWKME&  (i) & $-0.491$ ($ 0.194$) & $-0.127$ ($ 0.161$) & $ 0.135$ ($ 0.181$) & $ 0.729$ ($ 0.016$) &$ 0.197$ ($ 0.079$)\\
    &SAIWKME&  (i) & $-0.496$ ($ 0.186$) & $-0.133$ ($ 0.164$) & $ 0.129$ ($ 0.170$) & $ 0.729$ ($ 0.016$) &$ 0.198$ ($ 0.076$)\\
    \cmidrule{2-8}
    &\textbf{SIWKME-IV}&  (ii) & $-0.013$ ($ 0.339$) & $-0.056$ ($ 0.208$) & $ 0.099$ ($ 0.219$) & $ 0.737$ ($ 0.012$) &$ 0.138$ ($ 0.068$)\\
    &\textbf{SDRKME-IV}&  (ii) & $-0.031$ ($ 0.330$) & $-0.050$ ($ 0.206$) & $ 0.096$ ($ 0.210$) & $ \boldsymbol{0.738}$ ($ 0.012$) &$ \boldsymbol{0.134}$ ($ 0.068$)\\
    &SIWKME&  (ii) & $-0.484$ ($ 0.198$) & $-0.132$ ($ 0.168$) & $ 0.128$ ($ 0.179$) & $ 0.729$ ($ 0.016$) &$ 0.196$ ($ 0.078$)\\
    &SAIWKME&  (ii) & $-0.490$ ($ 0.189$) & $-0.138$ ($ 0.165$) & $ 0.121$ ($ 0.173$) & $ 0.729$ ($ 0.015$) &$ 0.196$ ($ 0.075$)\\
    
    \midrule \multirow{8}{*}{(c)}& \textbf{SIWKME-IV}& (i) & $ 0.014$ ($ 0.371$) & $-0.097$ ($ 0.268$) & $ 0.100$ ($ 0.225$) & $ 0.620$ ($ 0.014$) &$ 0.154$ ($ 0.079$)\\
    &\textbf{SDRKME-IV}& (i) & $-0.021$ ($ 0.364$) & $-0.094$ ($ 0.253$) & $ 0.092$ ($ 0.222$) & $ \boldsymbol{0.621}$ ($ 0.013$) &$ \boldsymbol{0.150}$ ($ 0.076$)\\
    &SIWKME& (i) & $-0.482$ ($ 0.225$) & $-0.132$ ($ 0.187$) & $ 0.149$ ($ 0.200$) & $ 0.614$ ($ 0.016$) &$ 0.200$ ($ 0.084$)\\
    &SAIWKME& (i) & $-0.485$ ($ 0.218$) & $-0.130$ ($ 0.184$) & $ 0.148$ ($ 0.196$) & $ 0.614$ ($ 0.015$) &$ 0.200$ ($ 0.081$)\\
    \cmidrule{2-8}
    &\textbf{SIWKME-IV}& (ii) & $-0.004$ ($ 0.376$) & $-0.090$ ($ 0.267$) & $ 0.116$ ($ 0.240$) & $ \boldsymbol{0.620}$ ($ 0.014$) &$ 0.157$ ($ 0.080$)\\
    &\textbf{SDRKME-IV}& (ii) & $-0.030$ ($ 0.366$) & $-0.084$ ($ 0.262$) & $ 0.111$ ($ 0.232$) & $ \boldsymbol{0.620}$ ($ 0.014$) &$ \boldsymbol{0.154}$ ($ 0.079$)\\
    &SIWKME& (ii) & $-0.478$ ($ 0.221$) & $-0.136$ ($ 0.200$) & $ 0.145$ ($ 0.201$) & $ 0.614$ ($ 0.015$) &$ 0.200$ ($ 0.082$)\\
    &SAIWKME& (ii) & $-0.482$ ($ 0.217$) & $-0.134$ ($ 0.191$) & $ 0.142$ ($ 0.193$) & $ 0.614$ ($ 0.015$) &$ 0.200$ ($ 0.082$)\\
    \bottomrule
\end{tabular}
\begin{tablenotes}
  \footnotesize
  \item \(\dagger\) \( S^*(t;\boldsymbol{\eta}^{opt}) \) are \(0.750\) and \(0.634\) for two settings, respectively.
\end{tablenotes}
    \label{tab7}
\end{table}

\section{Application to the cancer screening} \label{5}

Individuals aged 74 with a life expectancy of approximately 10 years should be assigned to sigmoidoscopy screenings based on their characteristics \cite[]{tangTimeBenefitColorectal2015}. However, \cite{tangTimeBenefitColorectal2015} did not provide specific regimes to guide individuals. Thus, our analysis includes a total of 17914 eligible subjects, with ages ranging between 70 and 78 at the time of trial entry, to explore how to assign screening for these subjects. These individuals possess complete information on relevant variables and have no prior history of any cancer, including colorectal cancer. Among them, \(8929\) \((49.85\%) \) participants were assigned to the control arm, and \(8985\) \((50.15 \%) \) participants were assigned to the intervention arm. Participants in the intervention arm received two colorectal cancer screenings,
while participants in the control arm did not receive any screenings. Even though the intervention was randomly assigned, non-adherence was observed among the participants. 
Of the \(17914\) participants,

\(2911\) \((16.25\%)\) participants did not adhere to the assignment. Thus, we utilize the randomized procedure as an IV to discover optimal treatment regimes. As we illustrated in Appendix \hyperref[app B]{A}, the randomized procedure can serve as a plausible IV if some important variables, mentioned in the following paragraph, are included. Survival time is measured as the interval from trial entry to death. Of the 5600 (31.26\%) subjects who were censored, 3533 (63.09\%) subjects were due to the end of the study, 1937 (34.59\%) subjects refused to answer, and 130 (2.32\%) subjects were censored for other reasons. Thus, it is more reasonable to posit a Cox proportional hazards model with some important variables in estimating the survival function of the censoring time \(S_C(s|Z, \boldsymbol{L}, A)\) rather than assuming that the administrative censoring occurs. 

Age (in years), sex (female denoted as 1), diabetes status (yes denoted as 1), family history of colorectal cancer (yes denoted as 1), and colorectal polyps (yes denoted as 1) are five significant confounders that can be used to predict the compliance type \cite[]{kianianCausalProportionalHazards2021,leeDoublyRobustNonparametric2021}. 
In contrast to \cite{leeDoublyRobustNonparametric2021} and \cite{kianianCausalProportionalHazards2021}, we have omitted the variable about the family history of any cancer because the findings in \cite{kianianCausalProportionalHazards2021} revealed that it does not act as a significant confounding factor. Additionally, the family history of colorectal cancer serves as an indicator of the family history of any cancer, potentially causing confusion in interpreting the parameters.
Denote the family history of colorectal cancer as colo\_fh and colorectal polyps as polyps\_f. Summary statistics of these confounders are reported in Table \ref{tab:realtab}. We employ t-tests for continuous variables and chi-squared tests for categorical variables to evaluate the balance of confounders between the groups designated by either the screening assignment (i.e., \(Z=1\), \(Z=0\)) or the actual screening status (i.e., \(A=1\), \(A=0\)). The \(p\)-values presented in Table \ref{tab:realtab} indicate that the comparison of the observed covariates with different values of \(Z\)  exhibit weaker evidence of systematic differences than the comparison of those with different values of \(A\). Nevertheless, we still observe some discrepancies between the two arms in diabetes. Thus, we further adjust the distribution of \(Z\) by positing a logistic model to estimate \(f(Z|\boldsymbol{L})\). Other nuisance models are posited as Section \ref{4.1}. The continuous variable, age, is min-max normalized to have minimum value \(0\) and maximum value \(1\). Our goal is to estimate the optimal treatment regimes from the linear regimes, with a view to maximizing the survival probability at the 1000th, 3000th, and 5000th days.

\begin{table}[!htbp]
  \setlength{\tabcolsep}{3pt}
\caption{\label{tab:realtab} Characteristics of the study population.}
  \centering
  \begin{tabular}{rllllll}
    \toprule
    Characteristics & Control ($Z=0$) & Intervention ($Z=1$) &  & Not screened ($A=0$) & Screened ($A=1$) &  \\ 
    & $n=8,929$ & $n=8,985$ &  & $n=11,840$ & $n=6,074$ &  \\ 
    \cmidrule(lr){2-4}  \cmidrule(lr){5-7}
   & \multicolumn{2}{l}{Number of Participants (\%)} & $P$-value & \multicolumn{2}{l}{Number of Participants (\%)} & $P$-value\\
    \midrule
    \multicolumn{4}{l}{\textbf{Age}\(^*\) (continuous)} \\
   & 71.688 (1.373) & 71.675 (1.372) & 0.524 & 71.702 (1.374) & 71.643 (1.370) & 0.007 \\ 
   \multicolumn{4}{l}{\textbf{Age Level}} \\ 
   70-73 yr & 7748 (86.773) & 7818 (87.012) &  & 10246 (86.537) & 5320 (87.586) &  \\ 
   74-78 yr & 1181 (13.227) & 1167 (12.988) & 0.653 & 1594 (13.563) & 754 (12.414) & 0.052 \\ 
    \multicolumn{4}{l}{\textbf{Sex}} \\
    Male & 4408 (49.367) & 4501 (50.095) &  & 5567 (47.019) & 3342 (55.021) &  \\ 
    Female & 4521 (50.633) & 4484 (49.905) & 0.338 & 6273 (52.981) & 2732 (44.979) & 0.000 \\ 
    \multicolumn{4}{l}{\textbf{Diabetes}}  \\
     No & 8066 (90.335) & 8022 (89.282) &  & 10628 (89.764) & 5460 (89.891) &  \\ 
    Yes & 863 (9.665) & 963 (10.718) & 0.021 & 1212 (10.236) & 614 (10.109) & 0.809 \\ 
   \multicolumn{4}{l}{\textbf{Family History of Colorectal Cancer}} \\  
  No & 7905 (88.532) & 7950 (88.481) &  & 10505 (88.725) & 5350 (88.080) &  \\ 
    Yes & 1024 (11.468) & 1035 (11.519) & \(0.934\) & 1335 (11.275) & 724 (11.920) & \(0.209\) \\ 
    \multicolumn{4}{l}{\textbf{Colorectal Polyps}} \\
   No & 8028 (89.909) & 8064 (89.750) &  & 10634 (89.814) & 5458 (89.858) &  \\ 
   Yes & 901 (10.091) & 921 (10.250) & 0.742 & 1206 (10.186) & 616 (10.142) & 0.947 \\ 
    \bottomrule
\end{tabular}
\begin{tablenotes}
  \footnotesize
  \item \(^*\) denotes a continuous variable, with the mean and the standard deviation reported.
\end{tablenotes}
\end{table}

The estimated optimal treatment regimes and their improvements compared to assignment all to \(1\) or \(0\) are presented in Table \ref{tabreal}. The 95\% confidence intervals for the difference between the estimated survival probabilities under the estimated optimal treatment regimes and simple regimes all to 1 or 0 are calculated using 500 bootstrap samples. The standard deviations given in the parentheses are calculated by the bootstrap samples as well. Across all settings, the left points of the intervals either stay above or near \(0\) when containing \(0\). The SIWKME-IV and the SDRKME-IV provide comparable estimates of the optimal treatment regimes. As the SDRKME-IV is better in simulation and theory, we recommend the estimated treatment regimes obtained by maximizing it. Based on the results of the SDRKME-IV, as \(t\) increases, an increasing number of adults are advised to screen: 4.706 percent of the population is recommended at \(t=1000\), 26.666 percent at \(t=3000\), and 92.994 percent at \(t=5000\). This exemplifies the immediate detrimental effects and long-term benefits of sigmoidoscopy screenings, as previously demonstrated in \cite{tangTimeBenefitColorectal2015}.

At \(t=1000\), only females with a family history of colorectal cancer will be advised due to their elevated risks of suffering from colorectal cancer.
At \(t=3000\), adults with risk factors for colorectal cancer, such as old age, diabetes, and a family history of colorectal cancer, are more likely to be recommended. 

By \(t=5000\), screening is generally recommended, with the exception of the elderly and individuals with colorectal polyps, who may face elevated risks from the screening process. {We also provide Table \ref{chars} to describe the characteristics of PLCO participants who are assigned cancer screening or not by the SDRKME-IV at different \(t\). At \(t=1000\), the recommended individuals are females aged 70 to 73 with a family history of colorectal cancer and no colorectal polyps. At \(t=3000\), the recommended individuals are more likely to be those aged 70-73, females, individuals with no diabetes, individuals with no family history of colorectal cancer, and individuals with no colorectal polyps. By \(t=5000\), the majority of individuals are recommended to screen. }

\begin{table}[!htbp]
\setlength{\tabcolsep}{3pt}
  \caption{The estimation results for the PLCO data.}
 
  \begin{tabular}{rrrrrrrrrrrrrr}
  \toprule
   \(t\) & Method &  Inter  & Age & Sex & Diabetes\_f &  Colo\_fh & Polyps\_f & \( \widehat{S}^*_{.}(t;\widehat{\boldsymbol{\eta}}^{opt}_{.}) \) & CI\(_1\times10^{-2}\) & CI\(_0\times10^{-2}\)  \\
   \midrule
   1000& {SIWKME-IV}& -0.639 & -0.370 &  0.205 & -0.149 &  0.592 & -0.201 & 0.968 (0.003) &  (-0.414, 1.018)  & (-0.027, 0.143)  \\
    1000& {SDRKME-IV}& -0.621 & -0.415 &  0.239 & -0.165 & 0.560 & -0.212 & 0.968 (0.003) & (-0.409, 1.021)  & (-0.022, 0.146)   \\

   3000& {SIWKME-IV}& -0.432 &  0.800 & 0.164 & 0.365 & 0.104 & -0.057 & 0.883 (0.006) & (0.003, 2.352)  & (0.170, 1.832) \\
   3000& {SDRKME-IV} & -0.491 & 0.639 & 0.285 & 0.463 & 0.230 & -0.048 & 0.882 (0.006) &(-0.030, 2.324)  &(0.173, 1.812)   \\

   5000& {SIWKME-IV}& 0.557 & -0.495 &  0.379 & -0.135 & -0.129 & -0.516 & 0.701 (0.006) & (-0.017, 0.853) & (-0.263, 3.597)  \\
   5000& SDRKME-IV &  0.506 & -0.476 & 0.250 & 0.268 & -0.413 & -0.462 & 0.700 (0.006) &(-0.143, 0.925) & (-0.184, 3.536)  \\
   \bottomrule
\end{tabular}

\begin{tablenotes}
  \footnotesize
  \item Note: {SIWKME-IV} refers to the estimator (\ref{S_I}), and {SDRKME-IV} refers to the doubly robust estimator (\ref{S_D}). The numbers in the parenthesis are the standard deviations. CI\(_1\) and CI\(_0\) denote the 95\% confidence intervals for the difference between the value functions calculated under the estimated optimal treatment regime and the simple treatment regime assigning all to treatments 1 and 0, respectively.
\end{tablenotes}
  \label{tabreal}
\end{table}

\begin{table}[!h]
\setlength{\tabcolsep}{3pt}
\caption{{Characteristics of PLCO participants assigned to cancer screening or not according to the SDRKME-IV. If the characteristic variable is categorical, the number of participants (\%) is reported, while for a continuous variable, the mean (the standard deviation) is reported. Trt1.1000 represents the group recommended for screening if \(t=1000\), and Trt0.1000 represents the group not recommended. Notations Trt1.3000, Trt0.3000, Trt1.5000, and Trt0.5000 follow the same logic.}} \label{chars}
\begin{tabular}{rrrrrrrr}
\toprule
Characteristics & Overall & Trt0.1000 & Trt1.1000 & Trt0.3000 & Trt1.3000 & Trt0.5000 & Trt1.5000 \\
 & (n=17914) & (n=17071) & (n=843) & (n=13139) & (n=4775) & (n=1255) & (n=16659) \\
    \midrule
\multicolumn{8}{l}{\textbf{Age}(continuous)}  \\
 & 71.682 (1.373) & 71.700 (1.382) & 71.302 (1.109) & 71.350 (1.239) & 72.594 (1.310) & 72.348 (1.173) & 71.632 (1.374) \\
\multicolumn{8}{l}{\textbf{Age Level}}  \\
70-73 yr & 15566 (86.893) & 14723 (86.246) & 843 (100.000) & 12183 (92.724) & 3383 (70.848) & 998 (79.522) & 14568 (87.448) \\
74-78 yr & 2348 (13.107) & 2348 (13.754) & 0 (0.000) & 956 (7.276) & 1392 (29.152) & 257 (20.478) & 2091 (12.552) \\
\multicolumn{8}{l}{\textbf{Sex}} \\
Male & 8909 (49.732) & 8909 (52.188) & 0 (0.000) & 7964 (60.613) & 945 (19.791) & 1125 (89.641) & 7784 (46.726) \\
Female & 9005 (50.268) & 8162 (47.812) & 843 (100.000) & 5175 (39.387) & 3830 (80.209) & 130 (10.359) & 8875 (53.274) \\
\multicolumn{8}{l}{\textbf{Diabetes}}  \\
No & 16088 (89.807) & 15267 (89.432) & 821 (97.390) & 12898 (98.166) & 3190 (66.806) & 1229 (97.928) & 14859 (89.195) \\
Yes & 1826 (10.193) & 1804 (10.568) & 22 (2.610) & 241 (1.834) & 1585 (33.194) & 26 (2.072) & 1800 (10.805) \\
\multicolumn{8}{l}{\textbf{Family History of Colorectal Cancer}}  \\
No & 15855 (88.506) & 15855 (92.877) & 0 (0.000) & 12412 (94.467) & 3443 (72.105) & 630 (50.199) & 15225 (91.392) \\
Yes & 2059 (11.494) & 1216 (7.123) & 843 (100.00) & 727 (5.533) & 1332 (27.895) & 625 (49.801) & 1434 (8.608) \\
\multicolumn{8}{l}{\textbf{Colorectal Polyps}}  \\
No & 16092 (89.829) & 15249 (89.327) & 843 (100) & 11654 (88.698) & 4438 (92.942) & 336 (26.773) & 15756 (94.580) \\
Yes & 1822 (10.171) & 1822 (10.673) & 0 (0.000) & 1485 (11.302) & 337 (7.058) & 919 (73.227) & 903 (5.420)\\
\bottomrule
\end{tabular}
\end{table}

We also calculate the estimated optimal treatment regimes by maximizing the SIWKME and the SAIWKME, which are reported in Table 5 of the supporting information for brevity. To compare the estimated treatment regimes on the PLCO data, we draw bootstrap samples over 500 times and estimate the 95\% confidence interval of the difference between the expected outcome following the estimated treatment regimes from SDRKME-IV and the others. The empirical expected outcomes of the estimated treatment regimes are evaluated with \(\widehat{S}_{SD}^*(t ;\boldsymbol{\eta})\). The overall results are presented in Table 6 of the supporting information. In summary, our estimators, the SIWKME-IV and the SDRKME-IV, perform similarly and have higher values compared to other methods. What stands out is at \(t=3000\), where the outcomes are \(0.8817\) and \(0.8819\) for the SIWKME-IV and the SDRKME-IV, respectively, while the outcomes are \(0.8728\) and \(0.8720\) for the SIWKME and the SAIWKNE, respectively. The confidence interval of the difference between the SDRKME-IV and the SIWKME is \((-0.1978, 2.0726)\times 10^{-2}\) and it is \((-0.2824, 2.0110)\times10^{-2}\) between the SDRKME-IV and the SAIWKME. At \(t=1000\) and \(t=5000\), the optimal treatment regimes estimated by all methods have no large difference from the treatment regime assigning all to treatment \(0\) and \(1\), respectively. Thus, their differences in outcomes are less significant than that at \(t=3000\). 

\section{Discussion} \label{discussion}
In this article, we establish the identification of the counterfactual survival function under any regime for a binary treatment subject to unmeasured confounding. In addition, we establish a doubly robust estimator to estimate the counterfactual survival function under any regime, which is consistent when one of the nuisance models is correct. The property of double robustness also allows us to take into account more clinically relevant information about the outcome model. Simulation has shown that our proposed approach has the desired performance. The data analysis has further clarified the improvements in the estimated optimal treatment regime. 

This paper studies a single-stage decision problem. However, it is worth noting that some diseases require sequential treatment over time. It would be of interest to extend our method to dynamic treatment regimes where a sequence of decision rules needs to be learned in the survival context with unmeasured confounding. Our optimization objective is the \(t\)-year survival probability, which can be generalized to the restricted mean \cite[]{gengOptimalTreatmentRegimes2015} and the quantile \cite[]{zhouTransformationInvariantLearningOptimal2022} of the survival time. We can also construct a multi-objective optimization framework to apply our method to the cancer screening data to trade off the survival probability and medical consumption. 
We here present the modified estimators with three different outcomes as follows:

\noindent{(a) The restricted mean survival time}
\[
    \widehat{R}_{\cdot}(t,\boldsymbol{\eta})=\sum_{i=0}^{n} I(\tilde{T}_{(i)}\le t)\widehat{S}_{\cdot}^{*}(\tilde{T}_{(i)};\boldsymbol{\eta})[\min
    \{\tilde{T}_{(i+1)},t\}-\tilde{T}_{(i)}],
\]

where $\tilde{T}_{(0)}=0, \{\tilde{T}_{(i)}\}_{i=1}^n$ are the order statistics of $\{\tilde{T}_i\}_{i=1}^n$ and $t$ is pre-determined time point. 

\noindent{(b) The $\tau$-th quantile survival time }
\[
    \widehat{Q}_{\cdot}^{*}(\tau,\boldsymbol{\eta})=\inf\{t:\widehat{S}^{*}_{\cdot}(t;\boldsymbol{\eta})\le 1-\tau\}, ~~\tau\in(0,1)
\]

\noindent{(c) Multi-objective combining survival probability and medical consumption}
\[ 
    \widehat{V}_{\cdot}(t,\eta)=\widehat{S}^{*}_{\cdot}(t ; \boldsymbol{\eta})-\lambda \mathbb{P}_{n} d_{\boldsymbol{\eta}}(\boldsymbol{L}),
\]
where \(\mathbb{P}_{n} d_{\boldsymbol{\eta}}(\boldsymbol{L})\) is the cost of treatments and \(\lambda\) is a hyperparameter used to trade off the survival probability and the medical consumption.

The asymptotic properties for the estimators are provided to analyze the convergence rate and show the advantage of the doubly robust estimator on both convergence rate and model misspecification. Consistency for the kernel smoothed estimators is also provided. However, the convergence rate for \(\widehat{{\boldsymbol{\eta}}}_{.}^{opt}\) is left to be solved for the very challenging it will be. The convergence rate for \(\widehat{{\boldsymbol{\eta}}}^{opt}_.\) might be slower than the classical \(\sqrt{n}\)-rate due to the indicator function \(I(\tilde{\boldsymbol{L}}^{T} {\boldsymbol{\eta}} \geq 0)\). It might be \(n^{1/3}\)-rate under a mild condition \cite[]{zhouTransformationInvariantLearningOptimal2022}. This is an interesting problem that deserves future research. The genetic algorithm is employed for optimization, which performs well with low dimensional \(\boldsymbol{L}\) \cite[]{kramer2017genetic}. When we have a large number of covariates, forward selection can be utilized to select important variables and overcome the difficulty of optimization \cite[]{zhang2018variable}. In this article, our focus is on studying the uninformative censoring assumption, which requires that censoring time is independent of survival time conditional on the observed covariates. If conditional independence requires an unobservable variable to be satisfied, exploring how to incorporate an IV to accommodate the unobserved variable is a future research direction that warrants thorough investigation.

\section*{Acknowledgement}
The authors thank the National Cancer Institute (NCI) for access to NCl's data collected by the prostate, lung, colorectal, and ovarian cancer screening trial. The statements contained herein are solely those of the authors and do not represent or imply concurrence or endorsement by NCI.

\section*{Supporting Information}
The supplementary simulations in Section \ref{4} can be found in the supporting information.
Our proposed method is implemented using R, and the R package \texttt{otrKM} to reproduce our results is available at \url{https://cran.r-project.org/web/packages/otrKM/index.html}. Access to the data used in the study is granted through the National Cancer Institutes and can be obtained upon approval through the following link: \url{https://cdas.cancer.gov/plco/}. Proofs for theorems in Section \ref{sec3} are available in the supporting information.

\section*{Appendix A: Illustration of the IV-related assumptions} \label{app B}
Recall that the IV-related assumptions are assumptions A2-A7 and Figure \ref{fig:causal graph} gives an illustration of these assumptions. Among these assumptions, assumptions A2-A5 are the core assumptions for all IV-based methods and assumptions A6-A7 are required for the identifiability of the final estimand \cite[]{wangIVEstimationCausal2022,cuiSemiparametricInstrumentalVariable2021,qiuOptimalIndividualizedDecision2021,wangBoundedEfficientMultiply2018}. Assumption A2 states that the variable \(U\) is sufficient to account for the unmeasured confounding. Assumption A3 requires that the IV is associated with the treatment conditional on \(\boldsymbol{L}\). Assumption A4 states that the direct causal effect from \(Z\) to \(T\) is all mediated by the treatment \(A\). Assumption A5 ensures that, conditional on \(\boldsymbol{L}\), the causal effect of \(Z\) on \(T\) is unconfounded by \(U\) because \(U\) has no causal effect on \(Z\).
Assumption A7 posits that in a model for the probability of being treated conditional on \(\boldsymbol{L}\) and \(U\), there exists no additive interaction between \(Z\) and \(U\). 
An example that satisfies this assumption is the additive probability model \(\pi(A=1|Z,\boldsymbol{L}, U)=g(Z,\boldsymbol{L})+ h(U,\boldsymbol{L})\), where \(g\) and \(h\) are measurable functions such that the range of the function \(\pi(A=1|Z,\boldsymbol{L}, U)\) is a subset of the interval \([0,1]\).
The additive probability model can capture certain types of real-world data \cite[]{caudillAdvantageLinearProbability}. Another interpretation of the assumption can be given as follows. Define the counterfactual treatment under the IV \(Z=z\) as \(A(z)\). In randomized trials where \(Z\! \perp \! \! \! \perp A(z)|(L, U)\) and \(Z\! \perp \! \! \! \perp A(z)|L\), or in observational studies under the same assumptions, the assumption A7 can be rearranged as \(E[A(1)-A(0)|\boldsymbol{L}, U]=E[A(1)-A(0)|\boldsymbol{L}]\).
Based on the relationship between the instrument \(Z\) and the treatment \(A(Z)\), we can segment the population into four adherence type: compliers \((A(1)=1, A(0)=0)\), always-takers \((A(1)=1, A(0)=1)\), never-takers \((A(1)=0, A(0)=0)\), and defiers \((A(1)=0, A(0)=1)\) \cite[]{imbensIdentificationEstimationLocal,angristIdentificationCausalEffects}. Then assumption A7 holds if the adherence type is determined by the observed covariates \(\boldsymbol{L}\), such as age and disease history. 


In our case, no unmeasured confounding assumption is problematic as health status is a confounder. However, the IV assumptions are more likely to be accepted. The IV relevance condition is convincing because the adherence rate is approximately 85\%. This indicates a strong relationship between the screening assignment (\(Z\)) and the treatment (\(A\)). Regarding the exclusion restriction assumption, the screening assignment itself does not directly affect the health of individuals. Instead, its effect is mediated by its influence on the treatment an individual receives, which ensures the exclusion restriction assumption. Since the IV we utilized here is the randomization procedure itself, it is independent of \(\boldsymbol{L}, U\) and each person has the same probability \(0.5\) of being assigned to the screening group. As a result, both the IV independence and IV positivity assumptions are satisfied. To achieve the independent adherence type assumption, it is necessary to include a list of variables that can predict the adherence type in our analysis. In our case, we use age, colorectal polyps, and diabetes to characterize the non-adherence due to health status. Other important variables are listed in Section \ref{5}.

{
\section*{Appendix B: Intuition underlying the IWKME-IV} \label{intuition}
We now provide more details on the intuition underlying our estimator (\ref{S_I}) of the counterfactual survival function. Our primary goal is to estimate the counterfactual survival function \(P(T^*(d_{\boldsymbol{\eta}}(\boldsymbol{L}))\geq t)\) via the Kaplan-Meier-like estimator as follows, where \(\prod_{s \leq t}\) is a product-based counterpart of the usual sum-based integral of calculus \cite[]{gill1990survey}. It requires a consistent estimator for the differential of the counterfactual cumulative distribution function \(dP(T^*(d_{\boldsymbol{\eta}}(\boldsymbol{L}))\leq s)\) and another estimator for the counterfactual survival function \(P(T^*(d_{\boldsymbol{\eta}}(\boldsymbol{L}))\geq s)\), where \(s\leq t\).
\[
    \widehat{S}^*_{KM}(t ;\boldsymbol{\eta})=\prod_{s \leq t}\left\{1-\frac{d P(T^*(d_{\boldsymbol{\eta}}(\boldsymbol{L}))\leq s)}{P(T^*(d_{\boldsymbol{\eta}}(\boldsymbol{L}))\geq s)}\right\}
\]

To be concise, we illustrate the intuition behind constructing the estimator (\ref{S_I}) with an example of constructing the estimator of the counterfactual survival function \(P(T^*(d_{\boldsymbol{\eta}}(\boldsymbol{L}))\geq s)\). Let \(d_{\boldsymbol{\eta}}(\boldsymbol{L})=1\), and for simplicity, omit the consideration of censoring and baseline covariates \(\boldsymbol{L}\). Therefore, our goal is to find a good estimator of ${P}(T^*(1)\geq s)$. Note that
\begin{align}\label{iden}
{P}(T^*(1)\geq s)=& E\left\{P(T^*(1)\geq s|U)\frac{\pi(A=1|Z=1,U)-\pi(A=1|Z=0,U)}{\pi(A=1|Z=1)-\pi(A=1|Z=0)}\right\}\notag\\
=&E\left\{P(T^*(1)\geq s|U)\frac{(2Z-1)\pi(A=1|Z,U)}{f(Z|U)(\pi(A=1|Z=1)-\pi(A=1|Z=0))}\right\}\notag\\
=&E\left\{P(T^*(1)\geq s|U)\frac{(2Z-1)I\{A=1\}}{f(Z|U)(\pi(A=1|Z=1)-\pi(A=1|Z=0))}\right\}\notag\\
=&E\left\{P(T^*(1)\geq s|Z,A,U)\frac{(2Z-1)I\{A=1\}}{f(Z)(\pi(A=1|Z=1)-\pi(A=1|Z=0))}\right\}\notag\\
=&E\left\{\frac{I\{T^*(1)\geq s\}(2Z-1)I\{A=1\}}{f(Z)(\pi(A=1|Z=1)-\pi(A=1|Z=0))}\right\}\notag\\
=&E\left\{\frac{I\{T\geq s\}(2Z-1)I\{A=1\}}{f(Z)(\pi(A=1|Z=1)-\pi(A=1|Z=0))}\right\},
\end{align}
where the first equality follows the independent adherence type assumption A7, the second equality is from the law of total probability, and the fourth equality follows from assumptions A2, A4, and A5. 
With the identification equation (\ref{iden}) above, we can construct $\widehat{P}(T^*(1)\geq s)$ mentioned below as an estimator of ${P}(T^*(1)\geq s)$. It is the same as the denominator part of our estimator (\ref{S_I}) when there is no censoring, the baseline covariates \(\boldsymbol{L}\) are omitted, and \(d_{\boldsymbol{\eta}}(\boldsymbol{L})=1\).
\begin{align*}
    \widehat{P}(T^*(1)\geq s)
    =&\frac{1}{n}\sum_{i=1}^n \frac{I\{T_i\geq s\} I\{A_i=1\} (2Z_i-1)}{\widehat{f}(Z_i)(\widehat{\pi}(A_i=1|Z_i=1)-\widehat{\pi}(A_i=1|Z_i=0))}
\end{align*}
}
\bibliography{An_anti-confounding_method}
\end{document}